# Images of Embedded Jovian Planet Formation At A Wide Separation Around AB Aurigae

Thayne Currie[1,2,3], Kellen Lawson[4], Glenn Schneider[5], Wladimir Lyra[6], John Wisniewski[4], Carol Grady[3], Olivier Guyon[1,5,7], Motohide Tamura[7,8,9], Takayuki Kotani[7,8], Hajime Kawahara[10], Timothy Brandt[11], Taichi Uyama[12], Takayuki Muto[13], Ruobing Dong[14], Tomoyuki Kudo[1], Jun Hashimoto[8], Misato Fukagawa[8], Kevin Wagner[5,15], Julien Lozi[1], Jeffrey Chilcote[16], Taylor Tobin[16], Tyler Groff[17], Kimberly Ward-Duong[18], William Januszewski[18], Barnaby Norris[19], Peter Tuthill[19], Nienke van der Marel[20], Michael Sitko[21], Vincent Deo[1], Sebastien Vievard[1,7], Nemanja Jovanovic[22], Frantz Martinache[23], Nour Skaf[1]

[1] Subaru Telescope, National Astronomical Observatory of Japan, Hilo, HI, USA. [2] NASA-Ames Research Center, Moffett Field, CA, USA. [3] Eureka Scientific, Oakland, CA, USA. [4] Homer L. Dodge Department of Physics and Astronomy, University of Oklahoma, OK, USA. [5] Steward Observatory, The University of Arizona, Tucson, AZ. [6] Department of Physics and Astronomy, New Mexico State University, NM, USA. [7] Astrobiology Center, Tokyo, Japan. [8] National Astronomical Observatory of Japan, Tokyo, Japan. [9] Department of Astronomy, Graduate School of Science, The University of Tokyo, Tokyo, Japan. [10] Department of Earth and Planetary Sciences, University of Tokyo, Tokyo, Japan. [11] Department of Physics and Astronomy, University of California-Santa Barbara, CA, USA. [12] Infrared Processing and Analysis Center, California Institute of Technology, Pasadena, CA, USA. [13] Division of Liberal Arts, Kogakuin University, Tokyo, Japan. [14] Department of Astronomy, University of Victoria, Victoria, BC, Canada. [15] NASA Hubble Fellowship Program – Sagan Fellow. [16] Department of Physics, University of Notre Dame, Notre Dame, IN, USA. [17] NASA-Goddard Spaceflight Center, Greenbelt, MD, USA. [18] Space Telescope Science Institute, Baltimore, MD, USA. [19] Sydney Institute for Astronomy, School of Physics, University of Sydney, Australia. [20] Leiden Observatory, Leiden, Netherlands. [21] Space Science Institute, Boulder, CO, USA. [22] Department of Astronomy, California Institute of Technology, Pasadena, CA, USA. [23] Université Côte d'Azur, Observatoire de la Côte d'Azur, CNRS, Laboratoire Lagrange, Nice, France

## Abstract

**Direct images of protoplanets embedded in disks around infant stars provide the key to understanding the formation of gas giant planets like Jupiter. Using the Subaru Telescope and Hubble Space Telescope, we find evidence for a jovian protoplanet around AB Aurigae orbiting at a wide projected separation (~93 au), likely responsible for multiple planet-induced features in the disk. Its emission is reproducible as reprocessed radiation from an embedded protoplanet. We also identify two structures located at 430-580 au that are candidate sites of planet formation. These data reveal planet formation in the embedded phase and a protoplanet discovery at wide, > 50 au separations characteristic of most imaged exoplanets. With at least one clump-like protoplanet and multiple spiral arms, the AB Aur system may also provide the evidence for a long-considered alternative to the canonical model for Jupiter's formation: disk (gravitational) instability.**

Almost all of the ~5000 known indirectly detected exoplanets orbit their host stars on solar system scales (a < 30 au)[1]. The core accretion model, where a young gas giant forms by slowly building up a multi-Earth mass core and then rapidly accreting protoplanetary disk gas, accounts for gas giants like Jupiter and Saturn at these locations[2]. In contrast, *directly-imaged* exoplanets typically have wide, 50-300 au orbits and are over ~5 times more massive than Jupiter[3-6]. Disk conditions might not allow in-situ formation for many of these planets by core accretion. A plausible alternative model is *disk instability*: a violent and rapid process of gravitational collapse that is best suited for forming supermassive gas giant planets at ~100 au[7].

Direct images of still-forming protoplanets embedded in disks around infant stars can provide critical clues as to where and how jovian planets on all scales form. The first incontrovertible protoplanet detections, PDS 70 bc, revealed jovian planet formation around a near-solar mass star[8-9] at scales (~20-35 au) smaller than the orbits of most imaged planets. Evidence of planet formation at wider separations by disk instability includes spiral density waves and spatially extended protoplanets undergoing gravitational collapse[7]. Many disks show structure potentially connected to planets on wide separations, including spiral arms[10]. However, until now, none of these systems show a direct detection of protoplanets themselves. More fundamentally, the single incontrovertible protoplanetary system known so far, PDS 70 bc, probes jovian planet formation near the final stages of assembly, where young gas giants have accreted substantial protoplanetary disk material, clearing out a cavity heavily depleted in gas and/or dust[8-9]. Detections of embedded protoplanets in gas and dust rich disks would reveal planet formation during earlier, key stages of assembly.

**Results**

To search for protoplanets, we conducted high-contrast imaging observations of the young, benchmark disk-bearing star AB Aurigae (AB Aur). We primarily used the Subaru Coronagraphic Extreme Adaptive Optics project (SCExAO)[11] on the Subaru Telescope between 2016 and 2020, coupled with the Coronagraphic High-Resolution Imager and Spectrograph (CHARIS) covering the major near-infrared (IR) passbands[12]. Supporting ground-based data combine SCExAO or other platforms with instruments covering complementary optical/IR wavelengths. We also observed AB Aur with the Space Telescope Imaging Spectrograph (STIS) on the Hubble Space Telescope (HST) in visible light in early 2021 and reanalysed newly-reprocessed archival HST data for the star from STIS (1999) and the Near Infrared Camera and Multi-Object Spectrometer (NICMOS; 2007).

**Detection of AB Aur b**. SCExAO/CHARIS data identify a bright concentrated emission source at $\rho \sim 0.59$" (~93 au) nearly due-south of the star in nine different data sets spanning 4 years (Table 1). This source, hereafter referred to as AB Aur b, lies exterior to and is easily distinguishable from disk features detected at $\rho \sim 0.1$"—0.4" and spiral structure at much wider separations (Fig. 1). It is located interior to AB Aur's millimeter-resolved dust ring. Its position matches the predicted location of a protoplanet that could explain the AB Aur disk's CO gas spirals[13-14]. The CHARIS image bears a striking resemblance to gas surface density maps from simulations of jovian planet formation on wide separations by disk instability[7,15] (Fig. 1).

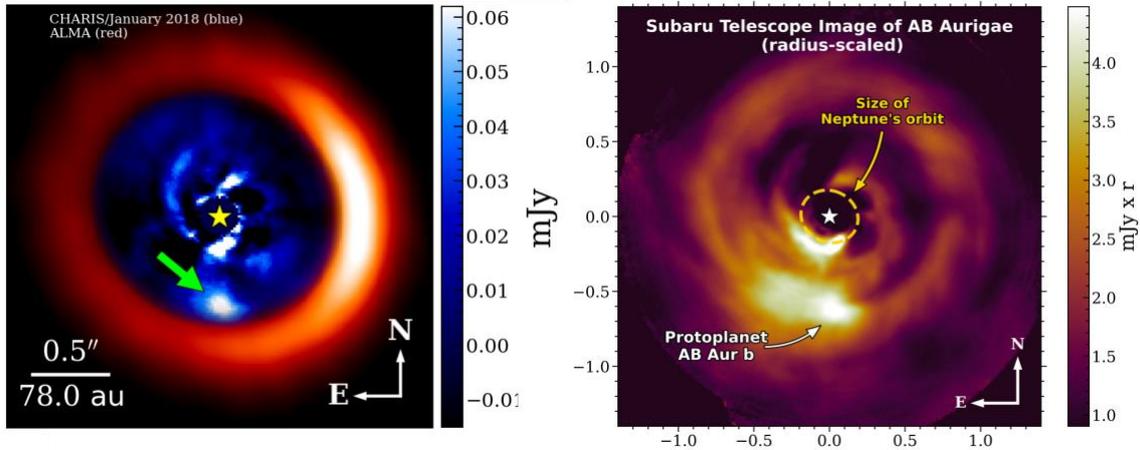

**Figure 1 – Detection of a clump-like protoplanet, AB Aur b, around AB Aur at wide separation (~93 au).** (left) SCExAO/CHARIS image from 6 January 2018 combined with ALMA submillimetre imaging[13] taken in 2014 at 900 $\mu m$ showing a ring of pebble-sized dust. AB Aur b lies interior to the dust ring at a location predicted from analysing spiral structure seen on smaller separations. The CHARIS component uses ADI/ALOCI for PSF subtraction. (right) Radius-scaled (i.e. image multiplied by separation *r*) CHARIS image of AB Aur from 2 October 2020 shown to highlight far fainter spiral structure in the disk. The green circle identifies AB Aur b. The CHARIS image uses polarimetry-constrained reference star subtraction to remove the stellar PSF. The color stretch is linear in this figure and all other figures.

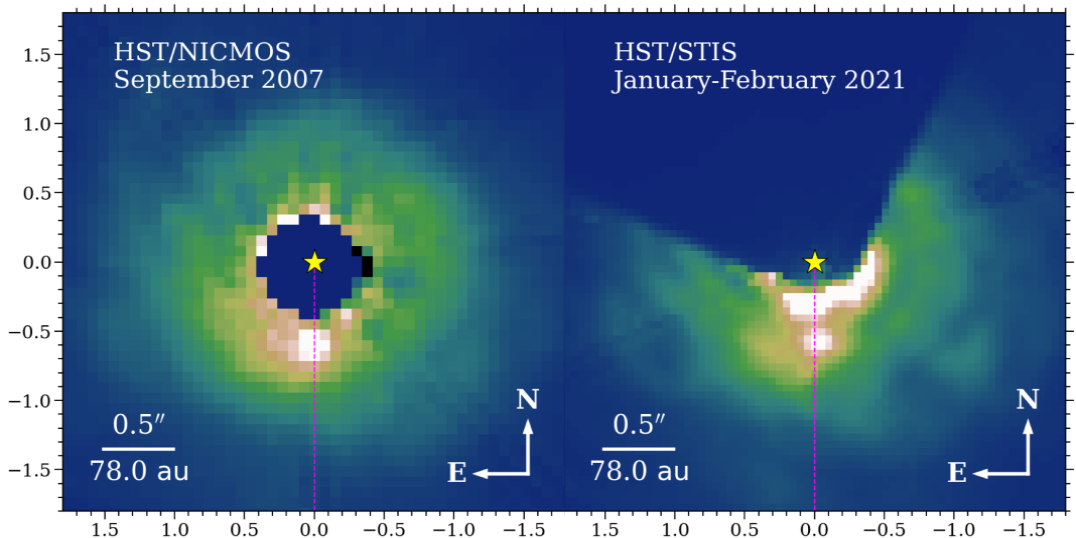

**Figure 2 – Hubble Space Telescope imaging of AB Aur over a 13-year time baseline:** NICMOS F110W data from 2007 (left) and STIS/50CCD from 2021 (right). In the NICMOS data, the coronagraph blocks central 0.3" region; the BAR5 occulter blocks the upper-left region of the STIS image. The dashed magenta line shows a position angle of 180°. In the 2007 (2021) data, AB Aur b lies to the left (right) of this line. The intensity scaling for both images is normalized to the peak count rate of AB Aur b: [-29,290] cts/s and [-23,230] cts/s for NICMOS and STIS, respectively. AB Aur b shows counterclockwise motion compared to the due-south position angle.

Furthermore, we recover AB Aur b in archival NICMOS total intensity data in the near-IR (1.1 μm) from 2007 and with STIS data acquired in early 2021 in visible light (0.58 μm) (Fig. 2). Visual inspection of the CHARIS, NICMOS, STIS data clearly shows that AB Aur b's position angle is changing, consistent with counterclockwise orbital motion. We easily rule out a stationary source at greater than a 5-$\sigma$ confidence level; a simple model of linear motion for AB Aur b's position with time yields a reduced chi squared close to 1.

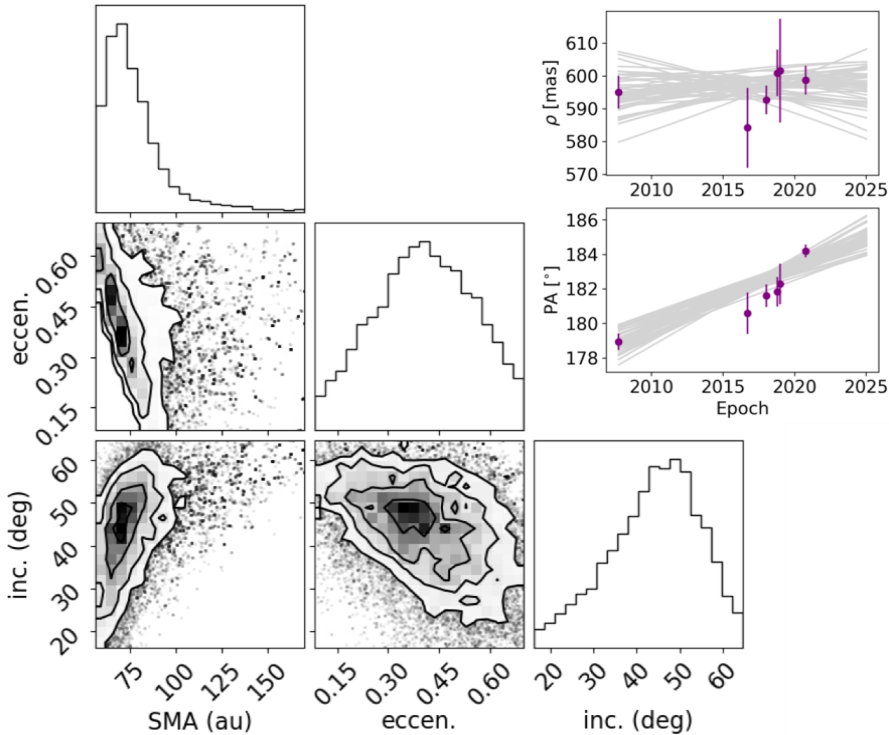

**Figure 3– Orbit fitting results for AB Aur b.** (left) Corner plot showing the posterior distribution. (inset) The angular separation and position angle for the 100 best-fitting orbits as a function of epoch. The astrometric error bars are 1-sigma error bars.

AB Aur b's morphology, location, and brightness compared to surrounding disk material cannot be due to processing artefacts. Point-spread function (PSF) subtraction using different approaches, that is, algorithms in combination with either angular differential imaging (ADI) or reference star differential imaging (RDI), recover AB Aur b with minimal evidence of photometric and astrometric biasing. Forward-modelling demonstrates that subtracting a synthetic source with properties like those we measure for AB Aur b entirely nulls the observed signal. See Methods and Supplementary Information Fig. 9 for more details.

**Orbits.** We constrained AB Aur b's orbit from its NICMOS and CHARIS astrometry using a well-tested Markov Chain Monte Carlo code[16] (Fig. 3). This modelling suggests that AB Aur b's orbit is viewed ~27-58º from face-on with an eccentricity of $e$ ~ 0.19-0.60, and a semi-major axis of ~45-143 au. Nearly all of these orbits do not cross the millimeter-resolved dust ring; a subset of them imply that AB Aur b is coplanar with the disk.

**Morphologies.** AB Aur b has a clearly defined center but the highest-quality data show it to be spatially resolved compared to a true point source. We estimate AB Aur b's size from multiple conservative reductions of the best data sets from CHARIS (January 2018; October 2020) and from HST/STIS and NICMOS, which use simple reference star subtraction largely immune from biasing. From CHARIS data, AB Aur b has an apparent radius of $\theta \sim 0.065$", or $\theta \sim 0.045$" (7 au) after deconvolving with CHARIS's intrinsic PSF; this is comparable to the Hill radius of 4 Jupiter-mass planet at 90 au. AB Aur b has a similar intrinsic radius in the range of $\theta \sim 0.053$"-0.073" in the more poorly-sampled STIS data.

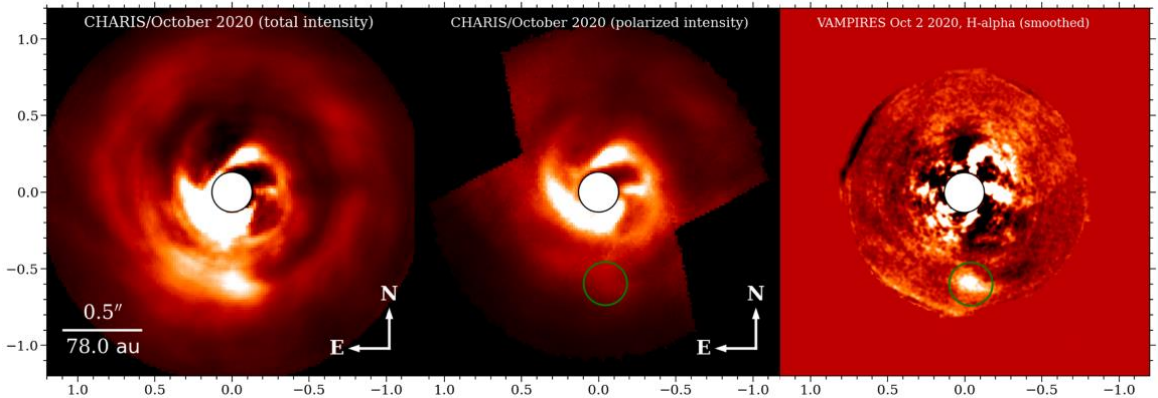

**Figure 4 – SCExAO/CHARIS images of AB Aur at different wavelengths and observing modes.** (left) CHARIS total intensity wavelength-collapsed image from October 2020 (same reduction as shown in Figure 1, right panel), showing a clear detection of AB Aur b. (middle) Polarized intensity wavelength-collapsed image obtained one day later. A pure scattered light disk feature would have been detected at the position of AB Aur b (green circle). Instead this region shows no concentrated emission indicating that AB Aur b is not detected. (right) emission at the approximate position of AB Aur b from VAMPIRES H$\alpha$ data using RDI/KLIP for PSF subtraction. From left to right, the intensity scaling is [0,0.0925] mJy, [0,0.055] mJy, and [-0.007,0.007] mJy, normalized to the source's apparent FWHM.

**Emission Sources.** AB Aur b cannot be a pure protoplanetary disk feature as its emission cannot be explained purely by scattered starlight (Fig. 4). At the same wavelengths where AB Aur b is detected in total intensity, previous and new polarized intensity imaging reveals complex disk structure but does not show concentrated emission consistent with AB Aur b[17]. The upper limit on AB Aur b's fractional polarized intensity in the near-IR is ~20%, lower than the 30% of surrounding disk material[18]. Visible Aperture Masking Polarimetric Imager for Resolved Exoplanetary Structures (VAMPIRES)[19] data reveal H$\alpha$ emission that may be attributed to AB Aur b. H$\alpha$ emission in excess of continuum emission is usually considered to be evidence of shocked, infalling gas onto a central object or accretion onto a protoplanet. For AB Aur b, most (perhaps all) of this H$\alpha$ signal instead may identify a scattered light component (Supplementary Section 5).

AB Aur b's near IR colors best resemble those of mid M to early L dwarfs (J—Ks = 1.14 +/- 0.28)[20], which are much redder than that of a bare stellar photosphere. However, note that the scattered `starlight' in the AB Aur disk originates not only from the 10,000K stellar photosphere (which dominates in the optical) but also from a 1,400 K component. The latter source is caused by sub-astronomical-unit scale gas and some dust emission and dominates in the infrared. Compared to the scaled spectrum of AB Aur originating from these two components (J—Ks = 1.71 +/- 0.02), AB Aur b is bluer in the near-IR. It is also underluminous with STIS in visible light (where it is detected) and with NIRC2 in the thermal IR (Lp; where it is not detected). AB Aur b's spectrum is consistent from epoch to epoch within errors and is bluer than that of the disk (see Supplementary Information Figures 10-12). AB Aur b has an apparent luminosity of $\log(L/L_\odot) \sim$ -2.60 to -2.79 (where $L_\odot$ is the solar luminosity), which is comparable to that of 1 Myr old, 10 Jupiter-mass planets[21-22]. Gaia and Hipparcos astrometry set a dynamical mass upper limit of ~0.13 $M_\odot$[23].

To reproduce AB Aur b's spectral energy distribution (SED), we considered a range of emission sources, including circumplanetary disks, bare (sub)stellar photospheres with various temperatures and gravities, and magnetospheric accretion models[24]. A dusty low-gravity atmosphere or smooth ~2,000-2,500 K blackbody combined with magnetospheric gas accretion reproduces all photometry and upper limits, and matches most of the CHARIS spectrum (see Supplementary Information Figure 12). In the best-fit composite model including the planet atmosphere, the planet has a mass of 9 Jupiter masses ($M_J$), a radius of 2.75 Jupiter radii ($R_J$), a temperature of 2,200 K, a surface gravity of $\log(g) = 3.5$, and is accreting at a rate of $\dot{M} \sim 1.1 \times 10^{-6}$ $M_J$ yr$^{-1}$. A bare photosphere does not reproduce the STIS photometry; an extended circumplanetary disk model predicts a NIRC2 detection instead of an upper limit. We emphasize that while AB Aur b's near-IR emission is well matched by a planet atmosphere, a simple blackbody of comparable temperature – e.g. from a circumplanetary envelope -- likewise reproduces this emission. See Supplementary Sections 2 and 8 for a detailed discussion.

**Detection of Additional Point-Like Features at Wider Separations.** The HST/STIS and NICMOS data identify two additional but far fainter concentrated emission sources at wider separations of $\rho \sim 2.75$" and 3.72" (429 au and 580 au). Over the nearly 22-year time interval between the STIS epochs, these sources appear in roughly the same location: they are not background objects lying behind and partially extincted by the disk. We identify no other similar sources in the STIS or NICMOS data. These sources are discussed in depth in Supplementary Section 4.

**Interpretation.** Despite the dissimilarity of AB Aur b's emission with scattered starlight, the SCExAO and HST images are not directly revealing only *thermal* emission from a planet atmosphere as is the case for the protoplanets PDS 70 b and c. AB Aur b is spatially resolved. A thermally-emitting object of its size (0.045" or ~7 au) and temperature (~2200 K) would have a luminosity far exceeding even that of the host star. Similarly, an object with AB Aur b's luminosity and size would have a temperature of ~23 K: far too low to emit detectable photons in the infrared.

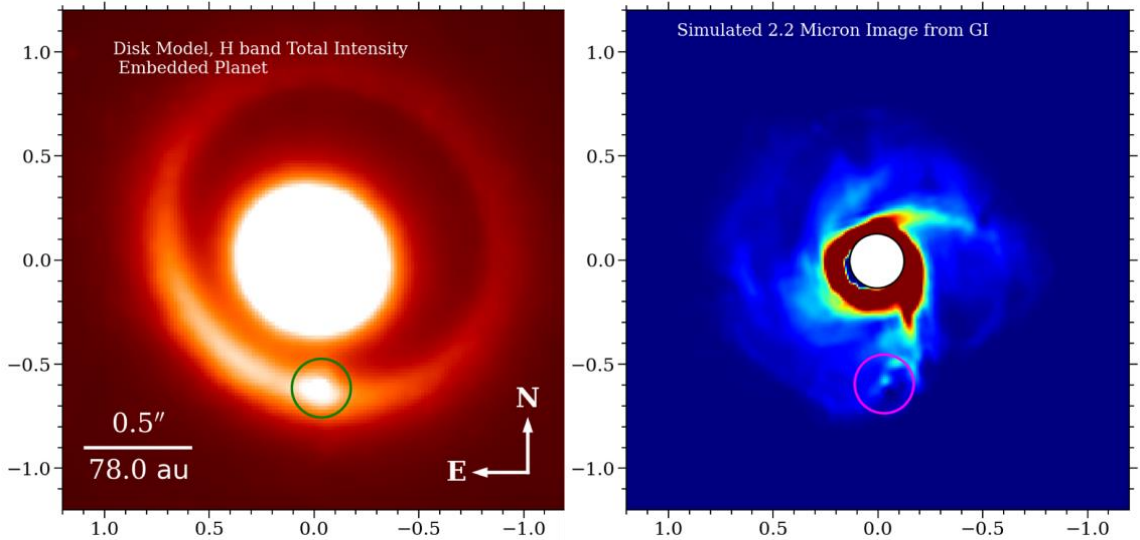

**Figure 5 – Simulating AB Aur b's emission and morphology.** (left) Synthetic image of AB Aur, its protoplanetary disk, and an embedded AB Aur b at 1.63 μm in total intensity (-0.01 to 0.041 mJy). (right) Synthetic image at 2.2 μm of a self-gravitating disk forming multiple clump-like protoplanets on scales comparable to AB Aur b in contrast units ([0,0.000125]). The resulting image shows a bright emission peak at the clump-like planet's location (circled). This is the densest, hottest clump in our GI simulation: other clumps created by GI are not visible because of their lower densities and/or projection effects of the system.

To explain AB Aur b's morphology and emission, we generated a synthetic image of scattered light from the AB Aur protoplanetary disk plus emission from an embedded protoplanet at a same location as AB Aur b. The protoplanet's emission is modelled as a simple blackbody of ~2200 K with a luminosity matched to AB Aur b's observed values. The model is agnostic as to whether this emission originates from a bare planet atmosphere or a circumplanetary envelope/disk. AB Aur b's embedded nature may favour the latter interpretation, and emission that does not originate from a planet atmosphere is needed to explain AB Aur b's optical detections. Our simulated embedded planet appears as a slightly extended object, in agreement with our data, because protoplanetary dust located nearby scatters light from the embedded planet (Fig. 5, left). The model planet photometry (1.4 mJy) roughly matches that measured for the protoplanet at H. The simulated image in polarized light shows no clear signal at the location of the planet, in agreement with our non-detection of AB Aur b in polarimetry (see Supplementary Fig. 17).

To explain AB Aur b within the context of planet formation models, we consider a disk instability model, where a self-gravitating disk produces multiple fragments at separations comparable to AB Aur b. From the resulting density and temperature profiles of the disk and clump as well as the disk scale height, we produce near-IR scattered light images of the system. In our scattered light image, the disk instability-formed clump produces a bright feature at the location of one clump, consistent with AB Aur b (Fig. 5, right).

**Discussion**

Evidence for at least one protoplanet around AB Aur at wide separations has important implications for our understanding of where planets form. While most studies analyze the demographics of imaged, fully-formed planets to constrain planet formation[25-26], the current location of an exoplanet may differ from where it formed. AB Aur b provides direct evidence that planets more massive than Jupiter can form at separations approaching 100 au, more than double the distance from the Sun to Kuiper belt objects like Pluto, and in striking contrast to expectations of planet formation by the canonical core accretion model.

      Finally, this discovery has profound consequences for our understanding of how planets form. AB Aur b provides a key direct look at protoplanets in the embedded stage. Thus, it probes an earlier stage of planet formation than the PDS 70 system. AB Aur's protoplanetary disk shows multiple spiral arms[17,27], and AB Aur b appears as a spatially resolved clump located in proximity to these arms. These features bear an uncanny resemblance to models of jovian planet formation by disk instability. AB Aur b may then provide the first direct evidence that jovian planets can form by disk instability. An observational anchor like the AB Aur system significantly informs the formulation of new disk instability models diagnosing the temperature, density and observability of protoplanets formed under varying conditions.

**Table 1| System Properties**

| Parameter | Measurement | References/Remarks |
|---|---|---|
| **AB Aur (star)** | | |
| Distance from the Sun | 155.9± 0.04 pc; 162.9±1.5 pc | Gaia eDR3 [28], Gaia DR2 [29] |
| Mass | 2.4± 0.2 $M_\odot$ | [30] |
| $T_{eff}$ | 9770 ± 100 K | Estimated from spectral type |
| Luminosity | 59 ± 5 $L_\odot$ | Estimated from radiative transfer modeling |
| Age (Myr) | 1—5 Myr | [30,31] |
| **AB Aur b** | | |
| Photometry (apparent magnitude) | $F_{cont}$(0.647 μm) = 1.13± 0.37 mJy<br>$F_{H\alpha}$(0.656 μm) = 3.01± 0.53 mJy<br>$m_{STIS}$ = 16.99 ±0.10<br>$m_{F110W}$ = 15.34± 0.12<br>$m_J$ = 15.15±0.19<br>$m_H$ = 14.56±0.18<br>$m_{Ks}$ = 14.01±0.21<br>$m_{Lp}$ > 13.34 (5$\sigma$) | Average of two epochs<br><br>JHK photometry from highest quality data set (20180106) |
| Epoch-Averaged Projected Separation | 0.592" ± 0.010"<br>(92.3 au ± 1.5 au) | Average of the NICMOS, STIS, and the highest-quality Ground-based Data |
| Astrometry (sep, PA°, date) | 0.591"±0.011", 178.9°±1.1°    20070914<br>0.584"±0.012", 180.6°±1.2°    20160918<br>0.593"±0.004", 181.6°±0.7°    20161015<br>0.601"±0.007", 181.8°±0.9°    20180106<br>0.602"±0.016", 182.3°±1.2°    20181016<br>0.599"±0.005", 184.2°±0.6°    20201002<br>0.575"±0.020", 182.5°±1.5°    20210104<br>                                                   20210204 | NICMOS, Highest-quality Ground-Based Data, and STIS |
| Apparent Luminosity | -2.695 ±0.095 Log($L_\odot$) | Estimated from SED modeling |
| Semimajor Axis | 93.9 au [44.6 au, 143.2 au] | Mean and 68% C.I. from orbit fit |
| Eccentricity | 0.40 [0.19,0.60] | Mean and 68% C.I. from orbit fit |
| Inclination (°) | 42.6° [27.1°, 58.2°] | Mean and 68% C.I. from orbit fit |
| Mass Estimates | < 130 Jupiter masses<br>10—12 Jupiter masses<br>9 Jupiter masses | Dynamical mass limit<br>Luminosity (1 Myr age)<br>SED modeling |

**Methods**

**Target Selection and Properties.** AB Aur – a 2.4 +/- 0.2 $M_\odot$, 155.9 pc distant member of the 1—3 Myr old Taurus-Auriga star-forming region[28-31] (Table 2) has long been considered to be a key laboratory for investigating planet formation. Optical to near-infrared (IR) scattered light imaging of its massive protoplanetary disk reveals numerous spiral arms on 200--500 au scales[32-34]. Millimeter imaging resolves a 170 au-wide ring of pebble-sized dust[15], while near-IR polarimetry tracing micrometer-sized dust reveals a highly structured disk interior to this ring[17,27]; these results are consistent with dust filtration from an unseen jovian protoplanet(s) inside the ring[35-36]. Analyses of CO gas spirals[14] suggest tidal disturbances from hitherto unseen protoplanets, including one at a position angle of ~180º and an angular separation of 0.6".

**Observations.** We observed AB Aur with the Subaru Telescope using SCExAO between 2016 and 2020, primarily coupled with the CHARIS integral field spectrograph in ``broadband", covering the major near-IR passbands simultaneously (JHK, $\lambda$ = 1.15—2.39 µm; R ~20). Except for our first (2016) epoch, we employed a Lyot coronagraph with a 0.13" radius occulting spot to suppress scattered starlight. We obtained one set of CHARIS data in a special integral field polarimetry mode also operating in ``broadband" to better distinguish between planet and disk signals[37]. Other SCExAO data utilized the HiCIAO camera in H band ($\lambda$ = 1.65 µm) coupled with the vector vortex coronagraph or the VAMPIRES instrument with data split between two optical bandpasses: a continuum bandpass (0.64 µm) and a 1nm–wide filter centered on H$\alpha$ (0.653 µm). We also acquired thermal IR ($\lambda$ = 3.78 µm) coronagraphic imaging with the Keck II Telescope using the NIRC2 camera. All observations were conducted in ADI mode[38], allowing the sky parallactic angle and thus off-axis astrophysical objects to rotate on the detector with time to better remove the noisy stellar halo.

Total on-source integration times and parallactic angle rotations varied substantially, covering rotations of 45—136º and exposure times of 20 minutes to 170 minutes. Weather conditions varied substantially between runs, ranging from extremely ``slow" 0.3" optical seeing to fast, variable 1.5" seeing. We obtained the eight total intensity CHARIS datasets. Our analysis focuses on the highest-quality observations, where the observing conditions and resulting raw image quality were by far the best (2016 Sept 9, 2018 Jan 6, 2018 Oct 16, and 2020 Oct 2), though poorer quality data do not identify any trends undermining our conclusions.

STIS coronagraphic imaging data from 2021 cover a broad single bandpass (0.2—1.0 µm) and utilized the BAR10 occulter[39] with the star placed in the corner region and consist of multiple visit-sets taken at different roll angles with interleaved observations of a color-matched PSF reference star (HD 21062). Archival 1999 STIS data place AB Aur behind the WedgeA1.0 position. In archival 2007 NICMOS data, the NIC2 camera coronagraph blocks light from the AB Aur primary: the system is observed in the F110W filter and the 2.0 µm linear polarizer and coronagraph. For archival data, we followed prior publications and identified color-matched reference PSF stars from the HST archive: HD 141653 for STIS and GJ 273 for NICMOS. Supplementary Information Section S1 clarifies some technical details unique to each observation.

**Basic Processing: Ground-based Data.** From raw CHARIS images, we extracted data cubes utilizing the standard CHARIS cube extraction pipeline[40]. Subsequent processing used the CHARIS Data Processing Pipeline[41]. Additional basic processing steps -- sky subtraction (if applicable), image registration, and spectrophotometric calibration -- followed previous steps. For the latter step, we used an empirical spectrum of AB Aur, obtained from the SpeX spectrograph and resampled from a resolution of R = 2,500 to R = 20. AB Aur's near-IR spectrum is far redder (J—K ~ 1.7) than the photosphere of an A0V star (J—K ~ 0) due to unresolved, au-scale hot circumstellar gas and dust[42]. Basic image processing for HiCIAO and NIRC2 data followed previous steps[43] including de-striping, bad pixel masking/correction, flat fielding, distortion correction, linearity correction and sky subtraction (for NIRC2 only), and precise image registration. For the VAMPIRES data, we dark subtracted and then shifted each 50ms sub-exposure to a common center, removing 10% of the sub-exposures with the poorest AO correction to improve the image quality of each of the coadded frames. Supplementary Information Section S1 describes additional technical details with our basic processing, especially with spectrophotometric calibration.

**Basic Processing: HST/STIS and NICMOS Data**. For both STIS and NICMOS data, we began our reductions with the HST pipeline-produced final data products (*sx2 files for STIS; *ima files for NICMOS). After cleaning the data of cosmic rays and hot/cold pixels, we determine absolute image registration. For STIS, we used the secondary spiders. For NICMOS, acquisition images were obtained in the F187N filter, where the PSF is not undersampled. Engineering telemetry information recorded in the science frames' fits headers listed the offset from the star's position in the unoccluded acquisition images to its position behind the coronagraph (the NOFFSETXP, NOFFSETYP keywords).

**PSF Subtraction: Ground-Based Data.** Advanced PSF subtraction techniques can attenuate disk and protoplanet signals in highly-structured disks like AB Aur's in difficult-to-model ways, greatly exacerbating the challenge of distinguishing between these two emission sources. Therefore, we adopted the following approach.

First, for all ground-based data sets with a suitable PSF reference star, we performed reference star differential imaging (RDI) using two complementary least-squares based methods: Karhunen-Lo'eve Image Projection (KLIP) algorithm and adaptive, locally-optimize combination of images algorithm (ALOCI)[44-46] applied over a single region of each image (i.e. ``full frame''). The nearby star HD 31233 was used as a reference PSF.

Second, we performed a full-frame, ``conservative'' implementation of KLIP and ALOCI in combination with ADI. For CHARIS data, we set the inner and outer boundaries for PSF subtraction at $\rho$ ~0.16'' and 1.05''. For HiCIAO and NIRC2 data, we set the inner radius to just beyond the occulting mask edges (at $\rho$ ~0.1'' and 0.25'', respectively) and outer radius to 1.05". For VAMPIRES, we set the inner radius to $\rho$ ~0.1" and outer radius to 0.8". For all data, we imposed a rotation gap criterion of $\delta >$ 1 PSF footprint to limit self-subtraction (i.e the subtraction of astrophysical signals by azimuthally displaced copies in the weighted PSF reference library). For KLIP, we retained only the first 1--5 KL modes; for ALOCI, we rewrote the covariance matrix using singular value decomposition (SVD) and truncated the diagonal terms at cutoffs of $10^{-2}$ to $10^{-6}$ before inversion.

Finally, we performed a ``polarimetry-constrained'' reference star differential imaging (PCRDI) for one of our highest quality data sets (CHARIS October 2020) as has been demonstrated before with CHARIS[36] and should result in near 100% throughput and no biasing of astrophysical sources. Briefly, we used the polarized intensity detection of the AB Aur disk obtained a day after this data set to conservatively estimate total intensity using the *diskmap* tool[47], assuming Rayleigh polarization with 100% peak fractional polarization and with disk inclination and PA of 30° and 60° respectively (from [29]). We then removed the estimate of the AB Aur total intensity disk signal in our entire observing sequence, leaving the residual signal predominantly composed of starlight from the stellar halo. Next, we performed a ``classical'' RDI reduction, identifying the linear combination of the PSF reference sequence that minimized the residuals with the disk subtracted AB Aur images. Finally, we subtracted the resulting cRDI PSF model from the original AB Aur CHARIS data (which includes the disk signal and AB Aur b) before derotating and combining the sequence as normal.

**PSF Subtraction: HST Data.** For HST data, the PSF is so stable that we can use simple, scaled reference star subtraction (cRDI) to suppress the stellar PSF. For both STIS and NICMOS, we used unsaturated acquisition images to determine the optimal relative flux scaling offset between AB Aur and its PSF reference star. While the diffraction spikes are not visible in the STIS acquisition images, we verified that the registered PSF reference star position in NICMOS data minimized diffraction spider residuals in both the unsaturated and coronagraphic images. The flux rescaling in STIS matches predictions from the reference star's R and I band photometry to within 1%. For the polarimetry data, we used POLARIZE to produce Stokes Q and U images, polarized intensity images, and total intensity images from each PSF subtracted image[48]. After performing PSF subtraction, we rotated each image north-up and created a master median-combined image (see S2 in Supplementary Information).

**Detection of AB Aur b.** Following standard, conservative practices, we define the statistical significance of this source's detection by comparing its integrated signal within an aperture to other aperture-summed pixels at the same separation and apply a finite-element correction[43,49]. Its detections are highly statistically significant (SNR = 5-12), even when we include real astrophysical signals (i.e. disk scattered light) in our noise estimate to determine the signal-to-noise ratio (SNR). In the highest-quality ground-based data (2018 January 06; 2020 October 02), AB Aur b has SNR ~10-12. In the poorest-quality ground-based CHARIS data set and the poorer-sensitivity VAMPIRES data, we detect AB Aur b at SNR ~5. Inspection of the reduced, sequence-combined CHARIS data cubes reveals AB Aur b in individual CHARIS spectral channels covering the J band channels through the blue half of H band (1.24--1.63 μm) for every CHARIS observation. In the highest-quality data, AB Aur b is visible in all CHARIS channels.

We recover AB Aur b in SCExAO/CHARIS, VAMPIRES, and HiCIAO data using either KLIP or ALOCI algorithms in combination with either ADI or RDI (where available). We detect it in polarimetry-constrained reductions in the October 2020 data with either cRDI or RDI-KLIP. AB Aur b has a consistent morphology, location, and brightness in each reduction of a given data set. Similarly, we detect AB Aur b in the 2021 HST/STIS and 2007 NICMOS total intensity data: AB Aur b is visible in PSF subtractions in individual roll angles with a consistent morphology, location, and brightness. AB Aur b is not detected in any dataset obtained in polarized light where only protoplanetary disk scattered light should be easily identifiable.

**Source astrometry, photometry and spectroscopy: ground-based data.** To derive precise astrometry for AB Aur b, we computed an initial centroid estimate from Gaussian fitting and then revised this estimate using a simple center-of-light algorithm applied to background-subtracted images. The background is defined from a region surrounding source pixels, offset in azimuth by 2--3 FWHM and radius by ~ 1 FWHM. Supplementary Information Section S3 describes how we determine astrometric uncertainties.

For the CHARIS data, satellite spots of a known brightness at ~16 $\lambda$/D separation established absolute flux calibration in each spectral channel. Instead of equating the aperture with a point-source FWHM we adopted an 8-pixel circular aperture $\theta$ ~ 0.13") for each channel. This aperture is comparable to the characteristic source size as measured by CHARIS; the measured encircled energy of the CHARIS PSF within 0.13" is nearly constant (~0.6) across CHARIS's bandpass for each data set. HiCIAO photometric calibration used neutral density filter observations of a zero-color star (HIP 32104) instead of satellite spots but adopted the same 0.13" aperture. The aperture used for photometry and spectral extraction was the same as that used for spectrophotometric calibration.

Forward-modeling measures and corrects for signal loss in our KLIP and ALOCI reductions employed for CHARIS and HiCIAO data and helps assess upper limits for NIRC2 data. Oversubtraction (confusion of the source and speckles) is the sole source of signal loss for the reductions utilizing RDI. For ADI reductions, signal loss can also be due to self-subtraction: annealing of the source in the target image by azimuthally offset copies in the reference image. For a model source intensity distribution, we adopted a simple Gaussian with a FWHM of 8 pixels, which well-resembles the real source intensity distribution. Forward-modeling yields channel-dependent throughputs between 90% and 99% for the RDI/KLIP and ADI/ALOCI reductions utilized in the four highest-quality CHARIS data sets: 18 September 2016, 06 January 2018, 16 October 2018, and 02 October 2020. HiCIAO and NIRC2 data reductions have comparably high throughput.

We extracted the AB Aur photometry and spectra from CHARIS, VAMPIRES, and HiCIAO data using the same aperture adopted for spectrophotometric calibration and subtracting same background used to constrain the source morphology. Spectra between different CHARIS epochs agree within errors. To verify our extracted source properties, we inserted a negative copy of the AB Aur b spectrum into registered, spectrophotometrically calibrated images and performed PSF subtraction with the same parameter settings used to detect the companion. AB Aur b's signal is nulled, leaving a flat background disk signal.

**Astrometry and Photometry: HST Data.** For NICMOS data, we derived an initial centroid for AB Aur b and the two wider-separation candidates estimate using Gaussian fitting. We then generated appropriate F110W PSFs using TinyTim[50] resampled at four times the native NICMOS pixel scale. Our adopted intrinsic source temperature is 1800 K, although differences with an A or G star template were minor, yielding an uncertainty of 5—10% in flux density. We subtracted a scaled, resampled TinyTim PSF model from the source location, varying both the brightness and position (+/- 1 pixels for each coordinate), until the background was flattened. We then calculated the total flux for an infinite aperture based off of the best-fit scaled and resampled TinyTim model. The NICMOS FITS headers provided the conversion from counts/s to Jy. For the better sampled STIS data, subtracting a simple Gaussian PSF equal to the measured source FWHM flattens the background: the fits header-provided conversion from to counts to Jy determined photometry for AB Aur b and wider-separation companions.

**Orbit Fitting.** We estimate AB Aur b's orbit from astrometry measured from NICMOS and CHARIS over a 13-year time baseline (2007 to 2020) using the sophisticated and widely used Markov Chain Monte Carlo code `orbitize!'[16]. We adopted priors of 6.413 +/- 0.1627 mas for the parallax to consider the different results from Gaia-eDR3 and Gaia DR2 and 2.4 +/- 0.2 solar masses for the primary mass. The Markov chains consisted of 40 temperatures for 1000 walkers with a burn-in of 500 steps and 1,000,000 total orbits sampled. As a check on our results, we performed additional fits using MCMC with different number of orbits/walkers or used the `Orbits for the Impatient' (OFTI) algorithm[18] in lieu of MCMC. These fits yielded nearly identical results.

We computed the mean and 68% confidence interval from the MCMC analysis for the primary mass, parallax, and orbital parameters of AB Aur b. In general, the MCMC analysis strongly favors an orbit where AB Aur b is viewed near aphelion. While the eccentricity is relatively high (mean eccentricity of ~0.41 and 68% C.I. of 0.25—0.56), very few of the best-fit orbits would cross the millimeter-resolved ring. The mean and 68% C.I. for the inclination are slightly higher than the literature quoted values for the inclination of the disk, although it is possible that AB Aur b is coplanar with the outer disk, given uncertainties. Due to the small fractional sampling of AB Aur b's orbit, some other orbital parameters – e.g. the longitude of the ascending node – are poorly constrained, although a subset of solutions do align with the disk position angle. Most of the orbits are consistent with the derived gas gap radius from the $^{13}$CO emission, estimated between 64 and 98 au[13]. The $^{13}$CO gap is well inside the dust cavity, just as seen in other transition disks, consistent with a gap cleared by a companion with the dust trapped at the outer edge of the gap[51].

**Spectroscopic Analysis.** To assess whether AB Aur b is compatible with pure protoplanetary disk scattered light, we compared its optical to mid IR SED to a scaled stellar spectrum. The optical portion of the spectrum is an A0V model reddened by Av=0.5 and matched to AB Aur's broadband photometry; the near-to-mid IR portion draws from the SpeX spectrum. The AB Aur b spectrum shows substantial differences with the model scaled stellar spectrum over all wavelengths. Its optical emission from STIS is underluminous compared to scattered starlight; its mid-IR upper limit from NIRC2 is a factor of 3 fainter than the predicted brightness for scattered light. The slope of the CHARIS spectrum is also bluer than scattered light. See Supplementary Information Figures 12-13.

**Modeling AB Aur b's Emission.** To compare AB Aur b to observational predictions for a simple embedded planet model, we generated spectra for combinations of several different emission sources: circumplanetary disk models spanning a range of planet masses (1—15 $M_J$), accretion rates ($10^{-5}$ to $10^{-8}$ $M_J$ yr$^{-1}$), disk inner radii (1—10 $R_J$), and magnetic field strengths (10—100 G)[21,24]. Instead of a planet atmosphere, we separately considered simple blackbody emission with temperatures comparable to that expected for young planets (~1,000—3,000K)[21].

To provide an explanation for the appearance of AB Aur b as an extended, embedded object, we used the Markov Carlo Radiative Transfer code MCMax3D to produce synthetic images of protoplanetary disk scattered light disk, thermal emission from AB Aur b[52], and light scattered by the disk from AB Aur b in the near-IR (where AB Aur b and the highly-structured disk are detected) and the millimetre (where only a small dust component near the star a ring of dust is detected). The embedded planet model adopts a source luminosity of $\log(L/L_\odot)$ ~ -2.60 and temperature of 2200 K embedded in AB Aur's protoplanetary disk. We adjust properties of the protoplanetary disk (e.g. scale height, dust mass) to match its characteristic near-IR scattered light brightness near the location of AB Aur b. To simultaneously match the millimetre ring imaged by the Atacama Large Millimeter Array (ALMA), we added a second disk component lying just exterior to AB Aur b. We convolved the model near-IR image with the CHARIS PSF. Because of the protoplanetary disk's low optical depth in our model ($\tau$ ~1.3 in H band), the model planet's signal is simply spread out over a large area, not significantly attenuated. The planet's intrinsic light may be further extincted by dust in a circumplanetary disk.

To link the appearance of the AB Aur system as a whole to models for planet formation, we generate a simple, proof-of-concept hydrodynamical disk instability model. The model consists of a self-gravitating two-dimensional global disk, ranging from 30 to 300 au, with gravitational unstable range between 50 and 150 au. To simulate planet formation by disk instability we use the PENCIL code[53]. The resolution is 432 x 864 in radius and azimuth, respectively. A few orbits into the simulation, the disk fragments into several (~10) clumps of the order of the mass of Jupiter, consistent with the Bonnor-Ebert mass of the system at 100 au (~0.97 $M_J$). To produce a synthetic image, we perform full radiative transfer post-processing with RADMC-3D[54]; we include the stellar blackbody, and choose one of the embedded clumps to also be a source of photons, with radius of two Jupiter radii and central temperature of 5,000 K.

Scattering is treated as isotropic. As the photons travel from the source to the atmosphere, they are successively scattered, absorbed, reemitted, and thermalized at a lower effective temperature.

**Data Availability**

With the exception of the first CHARIS epoch (obtained during engineering observations), all raw SCExAO data are available for public download from the Subaru SMOKA archive: https://smoka.nao.ac.jp/ . The first epoch data is available upon request. Keck data are available from the Keck Observatory Archive (https://koa.ipac.caltech.edu/cgi-bin/KOA/nph-KOAlogin); HST data are available from the Milkulski Archive for Space Telescopes (https://archive.stsci.edu/missions-and-data/hst ). Processed data are made available upon reasonable request.

**Code Availability.** Data reduction pipelines used to create CHARIS datacubes and perform subsequent processing are publicly available on Github (https://github.com/PrincetonUniversity/charis-dep; https://github.com/thaynecurrie/charis-dpp).

**Acknowledgements** We thank Anthony Boccaletti for many helpful conversations regarding the AB Aur protoplanetary disk and system properties. Zhaohuan Zhu generously provided circumplanetary disk models; Sarah Blunt provided expert advice on MCMC-based orbit fitting. We thank the Subaru, NASA-Keck, and Hubble Space Telescope Time Allocation committees for their generous allotment of observing time. This research is based in part on data collected at Subaru Telescope, which is operated by the National Astronomical Observatory of Japan. The authors acknowledge the very significant cultural role and reverence that the summit of Maunakea holds within the Hawaiian community. We are most fortunate to have the opportunity to conduct observations from this mountain. This paper makes use of the following ALMA data: 2012.1.00303.S. ALMA is a partnership of ESO (representing its member states), NSF (USA) and NINS (Japan), together with NRC (Canada) and NSC and ASIAA (Taiwan), in cooperation with the Republic of Chile. The Joint ALMA Observatory is operated by ESO, AUI/NRAO and NAOJ. This work was partially funded under NASA/XRP programs 80NSSC20K0252 and NNX17AF88G. The development of SCExAO was supported by the Japan Society for the Promotion of Science (Grant-in-Aid for Research #23340051, #26220704, #23103002, #19H00703 & #19H00695), the Astrobiology Center of the National Institutes of Natural Sciences, Japan, the Mt Cuba Foundation and the director's contingency fund at Subaru Telescope.

**Author contributions** T.C. conceived of the project, (co-)led the total intensity data reduction, performed the spectroscopic and orbital analysis, and wrote the manuscript. K.L. and J. W. led the polarized intensity data reduction and the polarimetry-constrained PSF subtraction method for CHARIS. G.S. planned the STIS observations and co-led the HST/STIS and NICMOS reductions. W.L. generated the hydrodynamical models used to compare the real data with models of planet formation. C.G. aided with project and observing planning. O.G., J.L., S.V. V. D., N.J., F.M., and N.S. oversaw the operation of SCExAO. M.T. provided project management. T. K and H. K. planned and obtained one epoch of CHARIS data. T.B. provided dynamical mass estimates. T. U. and B.N. contributed VAMPIRES data reduction steps. R.D., T. M. aided with interpreting planet-induced disk features. J.C., T.T., and T.G. lead the operation and maintenance of CHARIS. K. W.-D. and W. J. planned the STIS observations. NvdM provided the AB Aur ALMA image. M. S. obtained SpeX data. The authors contributed to the original observing proposals, data acquisition, and/or paper draft comments.

**Competing Interests** The authors declare no competing interests.

# Supplementary Information for

# Images of Embedded Jovian Planet Formation At Wide Separations Around AB Aurigae

Thayne Currie[1,2,3], Kellen Lawson[4], Glenn Schneider[5], Wladimir Lyra[6], John Wisniewski[4], Carol Grady[3], Olivier Guyon[1,5,7], Motohide Tamura[7,8,9], Takayuki Kotani[7,8], Hajime Kawahara[10], Timothy Brandt[11], Taichi Uyama[12], Takayuki Muto[13], Ruobing Dong[14], Tomoyuki Kudo[1], Jun Hashimoto[8], Misato Fukagawa[8], Kevin Wagner[5,15], Julien Lozi[1], Jeffrey Chilcote[16], Taylor Tobin[16], Tyler Groff[17], Kimberly Ward-Duong[18], William Januszewski[18], Barnaby Norris[19], Peter Tuthill[19], Nienke van der Marel[20], Michael Sitko[21], Vincent Deo[1], Sebastien Vievard[1,7], Nemanja Jovanovic[22], Frantz Martinache[23], Nour Skaf[1]

[1] Subaru Telescope, National Astronomical Observatory of Japan, Hilo, HI, USA. [2]NASA-Ames Research Center, Moffett Field, CA, USA. [3]Eureka Scientific, Oakland, CA, USA. [4] Homer L. Dodge Department of Physics and Astronomy, University of Oklahoma, OK, USA. [5]Steward Observatory, The University of Arizona, Tucson, AZ. [6]Department of Physics and Astronomy, New Mexico State University, NM, USA. [7]Astrobiology Center, Tokyo, Japan.[8]National Astronomical Observatory of Japan, Tokyo, Japan. [9]Department of Astronomy, Graduate School of Science, The University of Tokyo, Tokyo, Japan. [10]Department of Earth and Planetary Sciences, University of Tokyo, Tokyo, Japan. [11]Department of Physics and Astronomy, University of California-Santa Barbara, CA, USA. [12]Infrared Processing and Analysis Center, California Institute of Technology, Pasadena, CA, USA. [13]Division of Liberal Arts, Kogakuin University, Tokyo, Japan. [14]Department of Astronomy, University of Victoria, Victoria, BC, Canada. [15]NASA Hubble Fellowship Program – Sagan Fellow. [16]Department of Physics, University of Notre Dame, Notre Dame, IN, USA. [17]NASA-Goddard Spaceflight Center, Greenbelt, MD, USA. [18]Space Telescope Science Institute, Baltimore, MD, USA. [19]Sydney Institute for Astronomy, School of Physics, University of Sydney, Australia. [20] Leiden Observatory, Leiden, Netherlands. [21] Space Science Institute, Boulder, CO, USA. [22] Department of Astronomy, California Institute of Technology, Pasadena, CA, USA. [23] Université Côte d'Azur, Observatoire de la Côte d'Azur, CNRS, Laboratoire Lagrange, Nice, France. e-mail: currie@naoj.org

## S1. Robust Detection of AB Aur b

Supplementary Table 1 displays our full observing log. Below we add further details describing the detection of AB Aur b from ground based data and HST data.

**Supplementary Table 1 | Observing Log**

| UT Date | Instrument | Natural Seeing (") | Passband | $\lambda$ ($\mu m$) | $t_{exp}$ (s) | $N_{exp}$ | $\Delta$PA (°) | Observing/ Reduction Strategy |
|---|---|---|---|---|---|---|---|---|
| **New Data** | | | | | | | | |
| 20160918 | SCExAO/CHARIS | 0.3 | JHK | 1.16-2.39 | 45 | 60 | 60 | ADI |
| 20161015 | SCExAO/HiCIAO | 0.6 | JHK | 1.65 | 30 | 41 | 45 | ADI,RDI |
| 20180106 | SCExAO/CHARIS | 0.5 | JHK | 1.16-2.39 | 45 | 60 | 110 | ADI,RDI |
| 20180108 | SCExAO/CHARIS | 1.2 | JHK | 1.16-2.39 | 45 | 60 | 110 | ADI |
| 20180125 | SCExAO/CHARIS | 1.5 | JHK | 1.16-2.39 | 31 | 199 | 143 | ADI |
| 20181016 | SCExAO/CHARIS | 0.6 | JHK | 1.16-2.39 | 60.5 | 46 | 63 | ADI |
| 20181101 | Keck/NIRC2 | 0.7 | Lp | 3.78 | 50 | 120 | 132 | ADI |
| 20181221 | SCExAO/CHARIS | 0.9 | JHK | 1.16-2.39 | 20.6 | 162 | 53 | ADI |
| 20201002 | SCExAO/CHARIS | 0.4-0.7 | JHK | 1.16-2.39 | 31 | 130 | 127 | ADI,RDI |
| 20201002 | SCExAO/VAMPIRES | | H$\alpha$/Cont | 0.64-0.66 | 72.1 | 51 | 136 | ADI,RDI |
| 20201003 | SCExAO/CHARIS | 0.6-0.8 | JHK-pol | 1.16-2.39 | 60.5 | 73 | 80 | PDI |
| 20210104 | HST/STIS | - | 50CCD | 0.2--1 | 56 | 27 | 6 | RDI |
| 20210207 | HST/STIS | - | 50CCD | 0.2--1 | 56 | 18 | 4 | RDI |
| **Archival Data** | | | | | | | | |
| 19990123 | HST/STIS | - | 50CCD | 0.2--1 | 576 | 6 | 22.2 | RDI |
| 20071221- 20070914 | HST/NICMOS | - - | F110W | 1.10 | 511.6 | 2 | 137 | RDI |
| 20071221- 20070914 | HST/NICMOS | - - | POL0 120/240 | 2.05 | 192.0 | 18 | 137 | PDI,RDI |

**Ground-Based Near-IR Total Intensity Data**

Supplementary Figure 1 shows a gallery of ground-based images obtained with SCExAO/CHARIS and SCExAO/HiCiAO, reduced with ADI/ALOCI. For each image, the intensity scaling is normalized to the peak emission of AB Aur b. The detection of spiral features at small angular separations (~0.2") depends on the quality of a given data set. However, we clearly detect AB Aur b at $\rho \sim 0.6$" in each case.

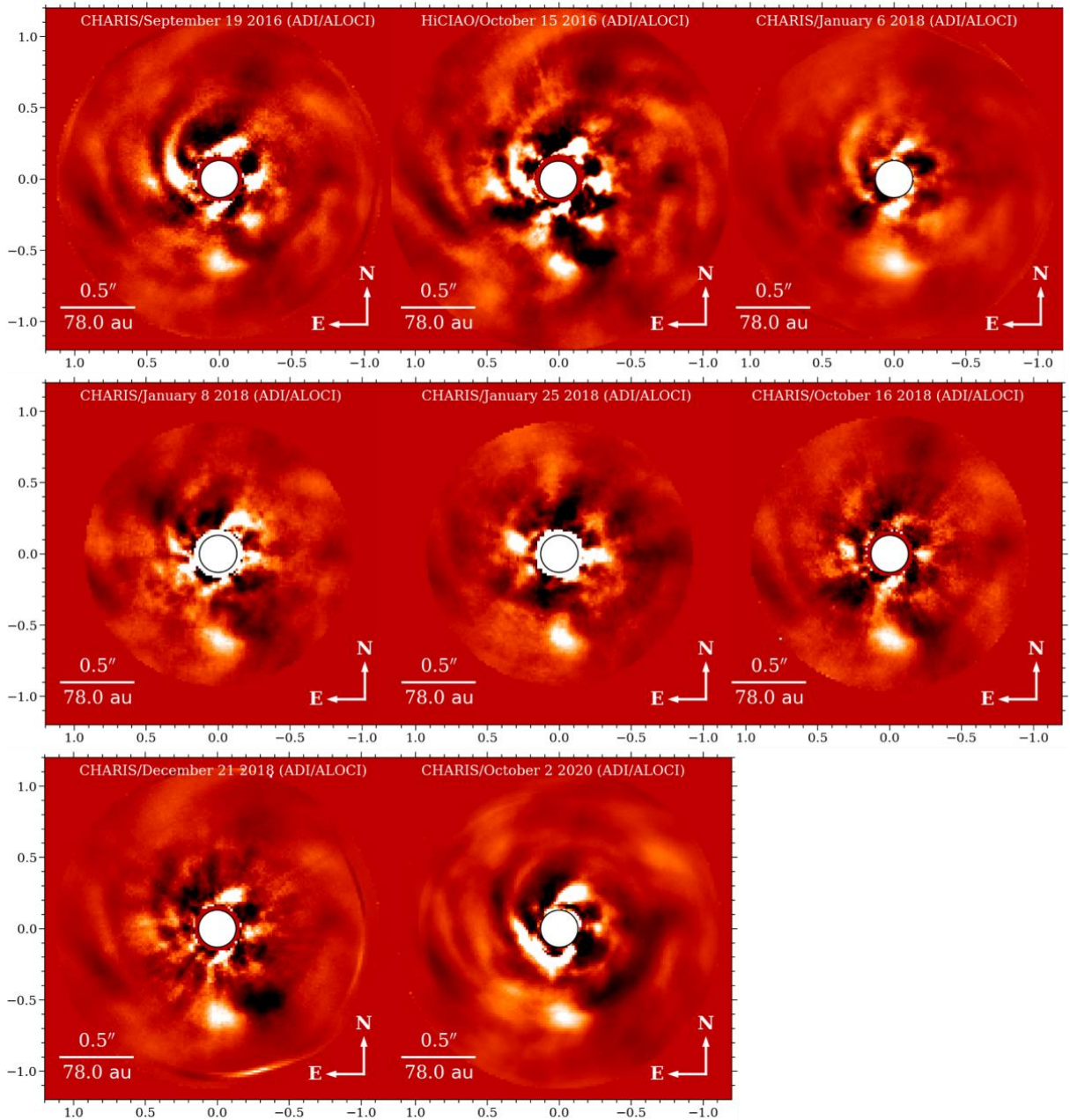

**Supplementary Figure 1 – Gallery of CHARIS and HiCIAO images of AB Aur** (ADI/ALOCI reductions). The data quality varies substantially from ``excellent'' (January 6, 2018; October 2, 2020) to "poor" (January 8 and 25, 2018), impeding our ability to consistently detect the inner spirals. However, AB Aur b is detected in all data sets. The panel intensity scales in units of mJy. Quantitatively, these values are (left to right, top to bottom) [-0.0375, 0.0375], [-0.004,0.004], [-0.065, 0.065], [-0.16, 0.16], [-0.18, 0.18], [-0.045, 0.045], [-0.0375, 0.0375], and [-0.0375, 0.0375]. The scaling for the two poorer-quality data sets is numerically larger because the flux calibration was done in a FWHM-sized aperture, not with a constant size.

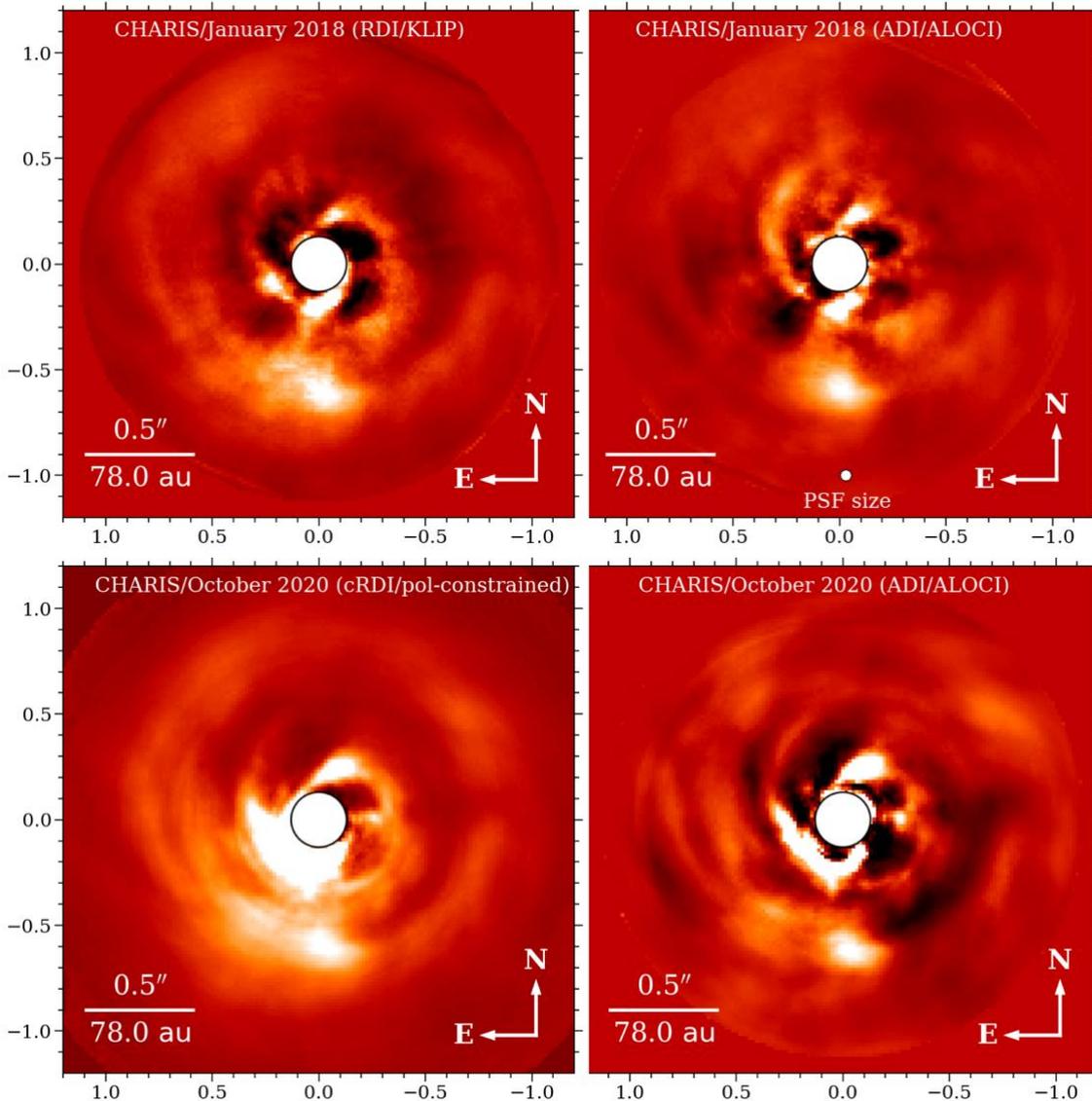

**Supplementary Figure 2 – CHARIS AB Aur data from January 2018 (top) and October 2020 (bottom) reduced using algorithms in combination with RDI (left panels) and ADI (right panels).** RDI reductions better preserve low surface brightness disk features, while ADI/ALOCI reductions better reveal spiral structure at the cost of self-subtraction at the smallest angular separations. The January 2018 ADI/ALOCI reduction is the same one shown in Figure 1 (left panel); the October 2020 cRDI/pol-constrained reduction is the same one shown in Figure 1 (right panel). All reductions clearly show AB Aur b. The filled white circle in the top-right panel shows the approximate PSF core size: AB Aur b is spatially resolved.

AB Aur b is visible and morphologically consistent in each SCExAO image regardless of whether we use ADI or RDI, ALOCI or KLIP (Figures S2 and S3). Generally speaking, ADI-ALOCI and ADI-KLIP yield higher SNR detections but act as better high-pass filters, significantly attenuating relatively flat emission from the disk, and sharpening their morphologies. PSF subtraction used in combination with RDI better preserve emission from the disk and in general yields a higher fidelity image of the astrophysical scene at the cost of more poorly suppressing speckle noise, especially at angular separations interior to AB Aur b.

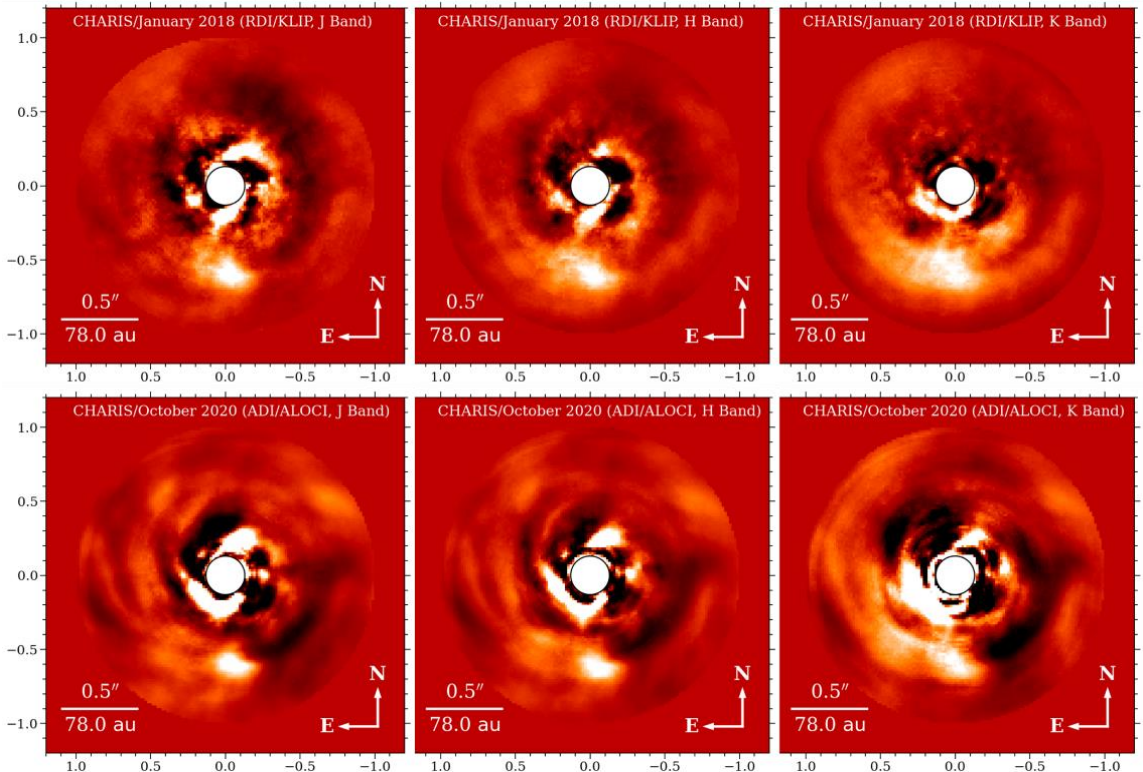

**Supplementary Figure 3 – The final J, H, and K images from CHARIS AB Aur data from January 2018 reduced with RDI/KLIP (top) and October 2020 reduced with ADI/ALOCI (bottom).** AB Aur b is plainly visible in J and H bands. At Ks band, its signal starts to become contaminated by the the disk.

Supplementary Figure 3 shows how the detectability of AB Aur b varies across J, H, and Ks passbands in the January 2018 RDI/KLIP and October 2020 ADI/ALOCI reduced data. At J band, AB Aur b's signal stands out clearly, its peak signal is roughly 4--5 brighter than the per-pixel average signal flanking disk regions in the RDI-reduced data. It is still clearly visible in H at ~2.5 times the disk brightness, although the disk background signal is larger in relative brightness. At Ks band, AB Aur b becomes contaminated by and is more difficult to distinguish from the disk. Only ADI reductions yield a clear signal in the longest wavelength channels in both data sets. These results are consistent with AB Aur b being an object with bluer near-IR colors than those expected from scattered starlight, whose SED resembles that of a ~1400 K blackbody[1].

We fail to decisively detect AB Aur b in the thermal IR (Lp ) with Keck/NIRC2 (Supplementary Figure 4). These data were taken with a poorer AO correction than that from most SCExAO data sets, although we do convincingly detect the ring of emission seen in submm data as well as the inner spirals; high humidity (near ~80%) also reduced thermal IR sensitivity at wide separations. A small subset of alternate, aggressive ADI reductions (not shown) do show a concentrated emission clump similar in spatial extent and location to b albeit at marginal significance (~$2\sigma$). Using more conservative reductions (shown) that better preserve disk features seen with other instruments leaves a non-detection (dashed circle). Likewise, it is not recovered in an RDI reduction. We also reduced other AB Aur Lp data sets from the Keck Observatory Archive, also failing to yield a detection.

Thus, we consider only 5-$\sigma$ upper limits for the detection of AB Aur b at Lp. Future thermal IR observations with the recently-upgraded Keck II AO system or with the James Webb Space Telescope are needed to detect or place better limits on AB Aur b in the thermal IR.

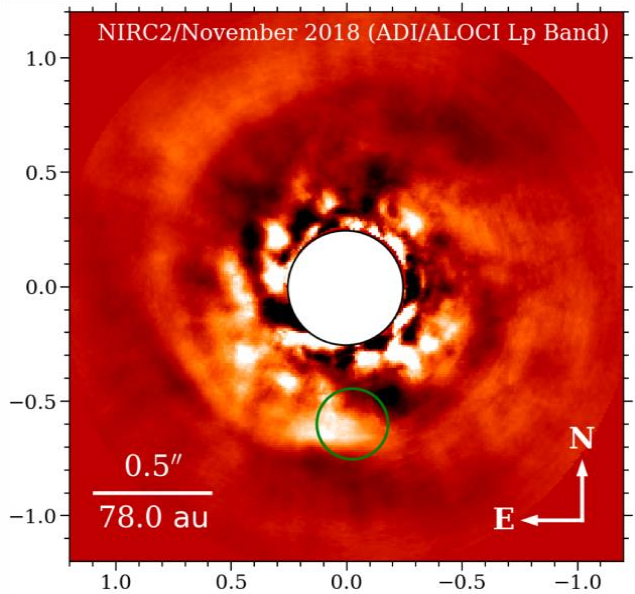

**Supplementary Figure 4 – Keck/NIRC2 Lp image of AB Aur.** A poorer AO correction (more variable, not diffraction limited), higher thermal background, and brighter disk explain to our failure to recover AB Aur b in these data (position circled).

### Ground-Based Optical Data

The VAMPIRES data were obtained in two cameras and split the continuum bandpass (0.64 μm; 2nm wide) and the narrow H$\alpha$ filter (0.653 μm; 1nm wide). We found that camera 1 yielded far more sensitive data, and thus we focus on these data. To precisely flux-calibrate these data, we used the empirical AB Aur spectrum[2] and computed the average flux density for wavelengths covering the continuum and narrow band filters. We estimate a flux density of ~5.2 Jy in the continuum and ~12.56 Jy at H$\alpha$ for the star.

As VAMPIRES' astrometric solution had not previously been calibrated, we constructed an approximate astrometric solution based off of recent observations of systems observed simultaneously with CHARIS: primarily based on binary companions at ~0.3" separations and the HD 1160 B brown dwarf. Based on these comparisons, we estimate a VAMPIRES pixel scale of 6.24 +/- 0.01 mas/pixel, yielding a field-of-view of ~0.8" in radius. We computed a north position angle offset from the detector y position of 78.6 +/- 1.2 degrees. The VAMPIRES detector likely has distortion on the order of ~a few pixels at the separation of AB Aur b, which we do not correct for: we have found no evidence so far that CHARIS suffers from astrometric distortion at a level that could impact our results.

We explored multiple approaches to reduce VAMPIRES data. Prior H$\alpha$ detections of protoplanet candidates either differenced ADI-reduced H$\alpha$ and continuum images or exploited high-resolution spectroscopy to detect the lines from SDI alone[3-4]. The VAMPIRES PSF was sufficiently stable to use HD 31233 as a PSF reference star and perform PSF subtraction using RDI with the KLIP algorithm. We also explored reducing VAMPIRES data with ADI in combination with the KLIP or ALOCI algorithms. Both RDI and ADI approaches yielded clear detections of AB Aur b (SNR ~5—5.8). Considering both intrinsic astrometric calibration uncertainties and uncertainties due to centroiding and the intrinsic source SNR, we derive a position of $\rho$ = 0.598" +/- 0.014", PA = 180.5º +/- 2.2º. While not nearly as precise as other measurements, this position is consistent with contemporaneous CHARIS astrometry at the 1.4-sigma level. The RDI reduction results in ~100% throughput and no self-subtraction. Thus, we base our VAMPIRES photometry on these data.

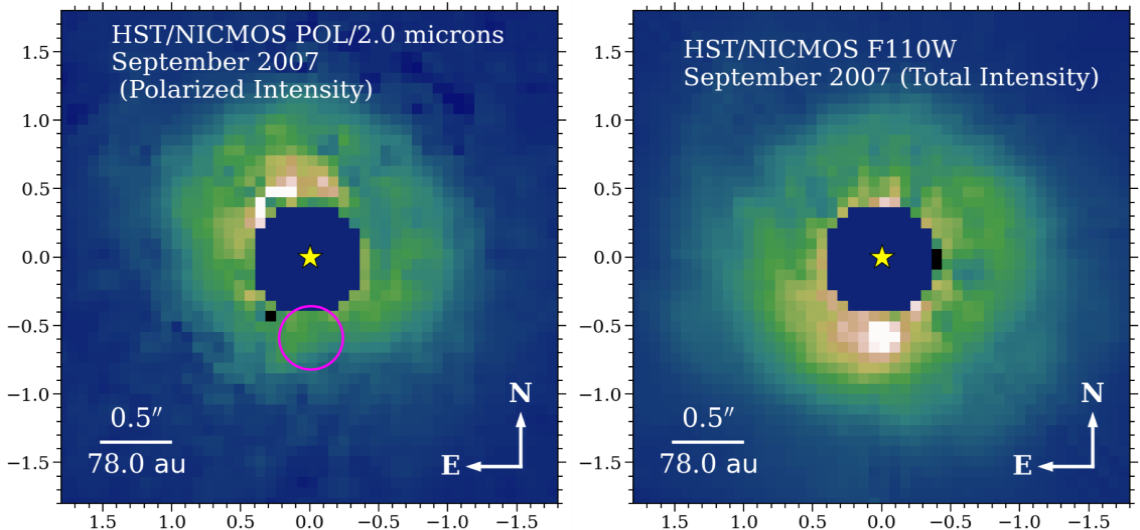

**Supplementary Figure 5 – HST/NICMOS 2.05 micron polarized intensity image for AB Aur. (left) compared to the F110W total intensity image (right).** We detect no polarized signal at the position of AB Aur b (circled).

### HST Total Intensity Data

AB Aur b is easily visible in the roll-combined F110W NICMOS data and is also separately detectable in the two individual rolls. The September 2007 NICMOS POL total intensity data (2.05 microns) also reveal what is potentially a detection of AB Aur b in the first roll position, albeit with substantial PSF subtraction residuals. However, the second, poorer quality roll position does not clearly reveal concentrated emission at this location. The roll-combined data therefore yields an ambiguous interpretation. Because the F110W NICMOS data yield an unambiguous detection, we focus on those data for photometry and astrometry.

Likewise, AB Aur b is easily visible in the roll-combined STIS images in both the January and February 2021 data sets, as well as individual PSF subtracted exposures. Comparing these data reveals some slight differences in the intensity distribution of the disk, suggesting possible variability. However, the signal of AB Aur b remains constant to within 10%. AB Aur b's position lies underneath the coronagraph wedge in the 1999 reprocessed archival STIS data.

**Additional Polarized Intensity Data**

As shown in Figure 4 of the main paper, AB Aur b is a non-detection in CHARIS polarized intensity imaging. Inspection of the total and polarized intensity images shows that the raw count values of pixels in total intensity have a definable peak at AB Aur b's position but are flat in polarized intensity. Supplementary Figure 5 likewise shows a non-detection of AB Aur b in HST/NICMOS polarimetry. Unpublished HiCIAO polarimetry of AB Aur likewise shows a detection of the disk but not AB Aur b.

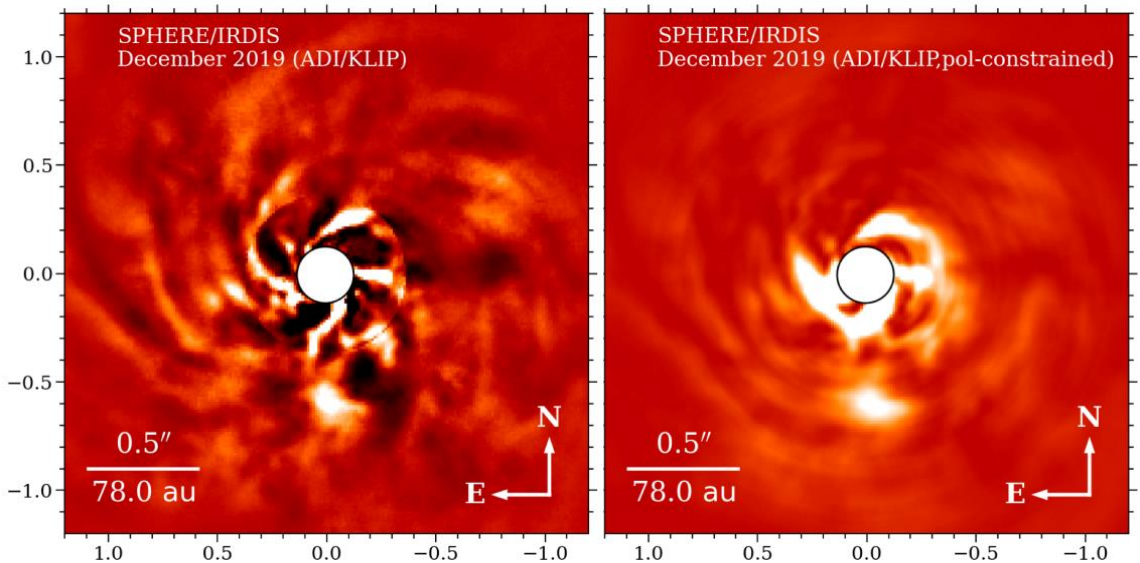

**Supplementary Figure 6 – Reprocessed SPHERE/IRDIS H band image of AB Aur from December 2019.** The data are reduced with ADI/KLIP (left) and ``polarimetry-constrained" ADI/KLIP (right). AB Aur b is visible in both reductions.

**Other Data Not Analyzed in this Work**

In addition to the data presented in this paper, we searched for additional AB Aur data from the ground and space that could affect our conclusions. Archival data from Subaru/CIAO from 2004 are of poorer quality compared to both SCExAO and Keck/NIRC2 data. We do detect a concentrated signal at roughly [E,N] = 0.05",-0.60", which would be consistent with AB Aur b in the CIAO data reduced with classical PSF subtraction, although the signal is close to the mask edge/saturated portion of the image. Thus, we do not yet consider this to be a decisive detection. Archival NIRC2 Lp data from 2015 and obtained with the vortex coronagraph do not appear to improve upon our 2018 NIRC2 limits.

As a final check on our results, we downloaded and reprocessed recently published SPHERE/IRDIS total intensity and polarized intensity data for AB Aur from December 2019. While the AO performance for these data is close in quality to our best SCExAO data sets, these data were obtained without a PSF reference star and with poor field rotation making them more ill-suited for unambiguous detections and characterizations of AB Aur b. We processed these data with the IRDAP pipeline[5]. To PSF subtract the total intensity data, we used two approaches: ADI/KLIP and polarimetry-constrained PSF subtraction method with KLIP.

As shown in Supplementary Figure 6, these data support conclusions based on SCExAO data. AB Aur b is clearly detected in SPHERE total intensity data. We do not detect AB Aur b in polarized intensity. The position of AB Aur b is intermediate between values determined from the October 2018 and October 2020 data. Results from these data are discussed further in S8.

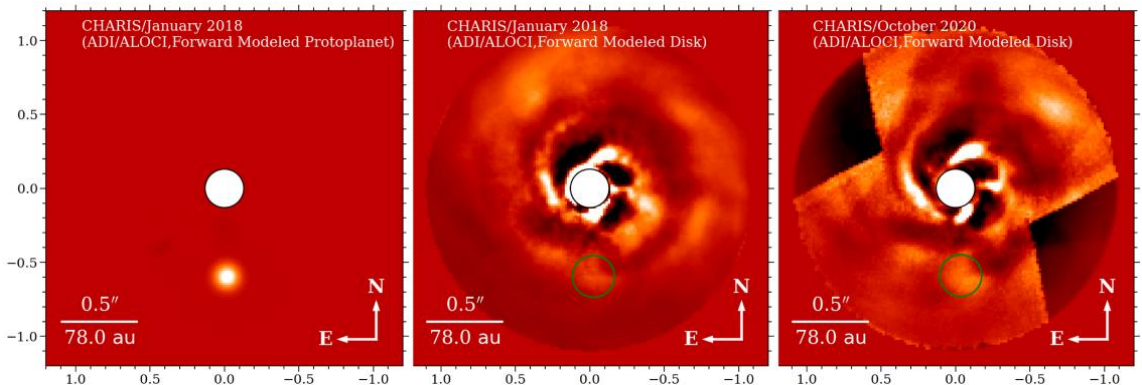

**Supplementary Figure 7 – Forward-modeling of our AB Aur images.** (left) Forward-model of a protoplanet through the January 2018 data: processing removes 10% of the light or less from this source. (middle) Forward-model of the Subaru/HiCIAO polarized intensity detection of the AB Aur disk through our CHARIS data; (right) forward-model of the SCExAO/CHARIS integral field polarimetry detection of the AB Aur disk through the CHARIS October 2020 data. The location of AB Aur b has a flat residual disk signal in both cases. Forward-modeling of the disk shows that our processing is not creating a point-like signal misidentified as a protoplanet. Note that the signals at the 10 o'clock and 4 o'clock positions from the AB Aur b positions are flat/constant, not concentrated. They are also at separations and position angles completely inconsistent with AB Aur b.

## S2. Forward-Modeling and Spectral Modeling: AB Aur b is not a processing artefact

To assess and correct for flux attenuation of AB Aur b in CHARIS data due to processing, we first performed forward-modeling with ADI-ALOCI and RDI-KLIP following previous demonstrated approaches with SCExAO data[6]. Supplementary Figure 7 (left) shows the forward-modeled PSF. The post-forward modelling signal from AB Aur b matches the input PSF, suggesting only minor reduction of its signal, consistent with the high throughputs (90-99%) listed in Methods. Similarly, we found negligible astrometric biasing due to processing in these data sets.

To assess how the source throughput depends on the assumed source morphology and intensity distribution, we performed forward-modeling on an unresolved point source, an 8-pixel wide (or 16 pixel-wide for HiCIAO) constant intensity source ($\theta \sim 0.13$"), and other intensity distributions. In all cases, signal loss at AB Aur b's location is negligible, less than 10%. Significant spectrophotometric and astrometric biasing only occurs for more aggressive reductions: e.g. PSF subtraction performed in small annular wedges instead of over the entire field at once or smaller rotation gaps.

Advanced PSF subtraction methods have the potential to break up continuous disk features to make them appears like planets[7]. To test this possibility with our SCExAO/CHARIS data, we adopted the polarimetry data as a disk model and then used forward-modeling techniques to simulate the appearance of these data (containing the polarized light detection of the disk) in the total intensity data. As shown in Supplementary Figure 7 (middle, right), AB Aur b cannot be a processing artefact: the forward-modeled disk image lacks a detectable concentrated signal at the companion's location. As a second test, we forward-modeled both the CHARIS and unpublished HiCIAO H band polarimetry data from 2016 through our January 2018 CHARIS data. These analysis yields identical results: the forward-modeled disk image shows no peaked signal at the position of AB Aur b.

**Optimized PCRDI** - Finally, we employ polarimetry-constrained classical reference star differential imaging (PCRDI) on the October 2020 data as an additional test. PCRDI is summarized in Methods. We give more detail below.

Let $I_{est}$ be the total intensity estimate determined from PI and a particular set of diskmap parameters describing a smooth scattering surface ( $h(r) = a+br^c$ ) with a particular peak fractional polarization (s), viewed at a particular orientation (incl, PA). Let $T_{D,est}$ be the sequence of $I_{est}$ rotated to the parallactic angles to match the target sequence, T, such that T - $T_{D,est}$ is our estimate of the target sequence containing only starlight.

Let M(T, R) be the PSF model for T based on the reference data R. For conventional RDI, the residuals would be: $T_{res}$ = T - M(T,R). For PI-constrained RDI (PCRDI), the residuals would be Tres = T - M(T-TD,est, R). In either case, M(T, R) could be a PSF model constructed with a least-squares algorithm like KLIP or with a simple linear combination of reference frames. In this case, we use the latter.

If we instead evaluate y = [T- $T_{D,est}$] - M(T- $T_{D,est}$, R) (i.e. subtracting the same PSF-model from the disk-subtracted target data), then the result is indicative of how well our estimate of the disk matched the true disk signal in the data -- being positive where we underestimated the true signal, and negative where we overestimated it. PCRDI can be optimized by seeking the values of (a, b, c, s, incl, PA) such that $\Sigma y^2$ is minimized. We perform the optimization y over a region excluding and thus unbiased by AB Aur b.

Supplementary Figure 8 below compares our best-fit estimate the protoplanetary disk scattered light– the polarized intensity divided by the best fit *diskmap* fractional

polarization – to the PCRDI-reduced total intensity image and the residuals within our training zone, y. The disk scattered light lacks the pronounced peak seen in the total intensity data that corresponds to AB aur b.

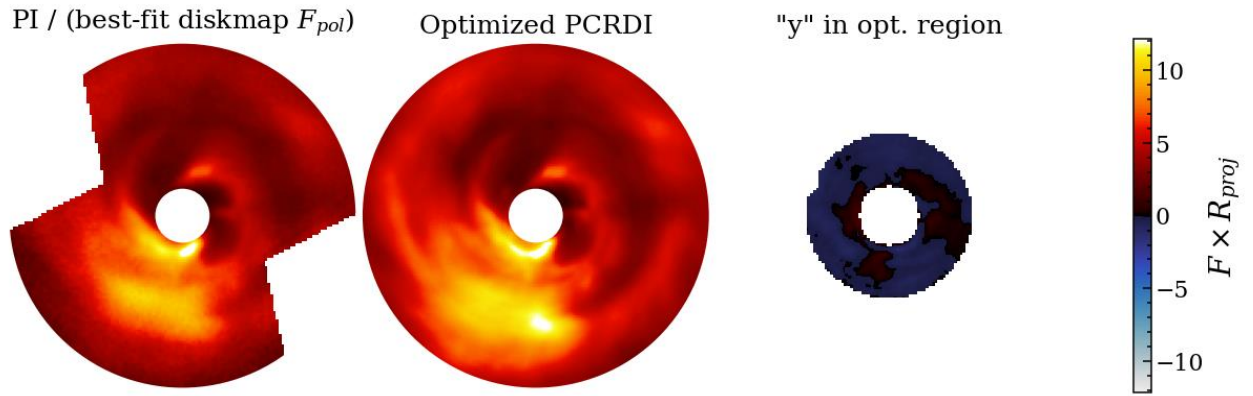

**Supplementary Figure 8** – Demonstration that AB Aur b is not a disk feature using PCRDI. (Left) CHARIS PI divided by the optimized diskmap fractional polarization estimate. This is the estimate of total intensity used to improve RDI in PCRDI. (Center) CHARIS optimized PCRDI result. (Right) The objective function for optimized PCRDI, i.e., y = [T-TD,est] - M(T-TD,est, R), which has been derotated and averaged over the sequence. Within the region considered, the total intensity estimate results in a final product that is effectively nulled. All images have been averaged over wavelengths and multiplied by the stellocentric separation assuming the best fitting diskmap scattering surface.

In summary, forward-modeling analyses of CHARIS and HiCIAO data rules out the possibility that AB Aur b is a disk feature artificially made to look like a concentrated source due to aggressive processing. Separately, the detection of AB Aur b using simple classical reference star PSF subtraction in STIS and NICMOS and polarimetry-constrained classical PSF subtraction with CHARIS supports the interpretation that AB Aur b is not a disk feature.

To assess the accuracy of our spectra, photometry, and estimated morphology of AB Aur b, we inserted a negative copy of our synthetic AB Aur b signal used in forward-modeling with a spectrum equal to that we extract from our real data (Supplementary Figure 9). As shown in Supplementary Figure 9 (top-right panel), this negative spectrum nulls the real AB Aur b signal, leaving a flat background. The bottom-right panel of Supplementary Figure 9 shows our HST/NICMOS image with the scaled, synthetic AB Aur b PSF removed. Here too, the residual image shows a flat background at the position of AB Aur b. Removing a PSF from the STIS data shows the same results.

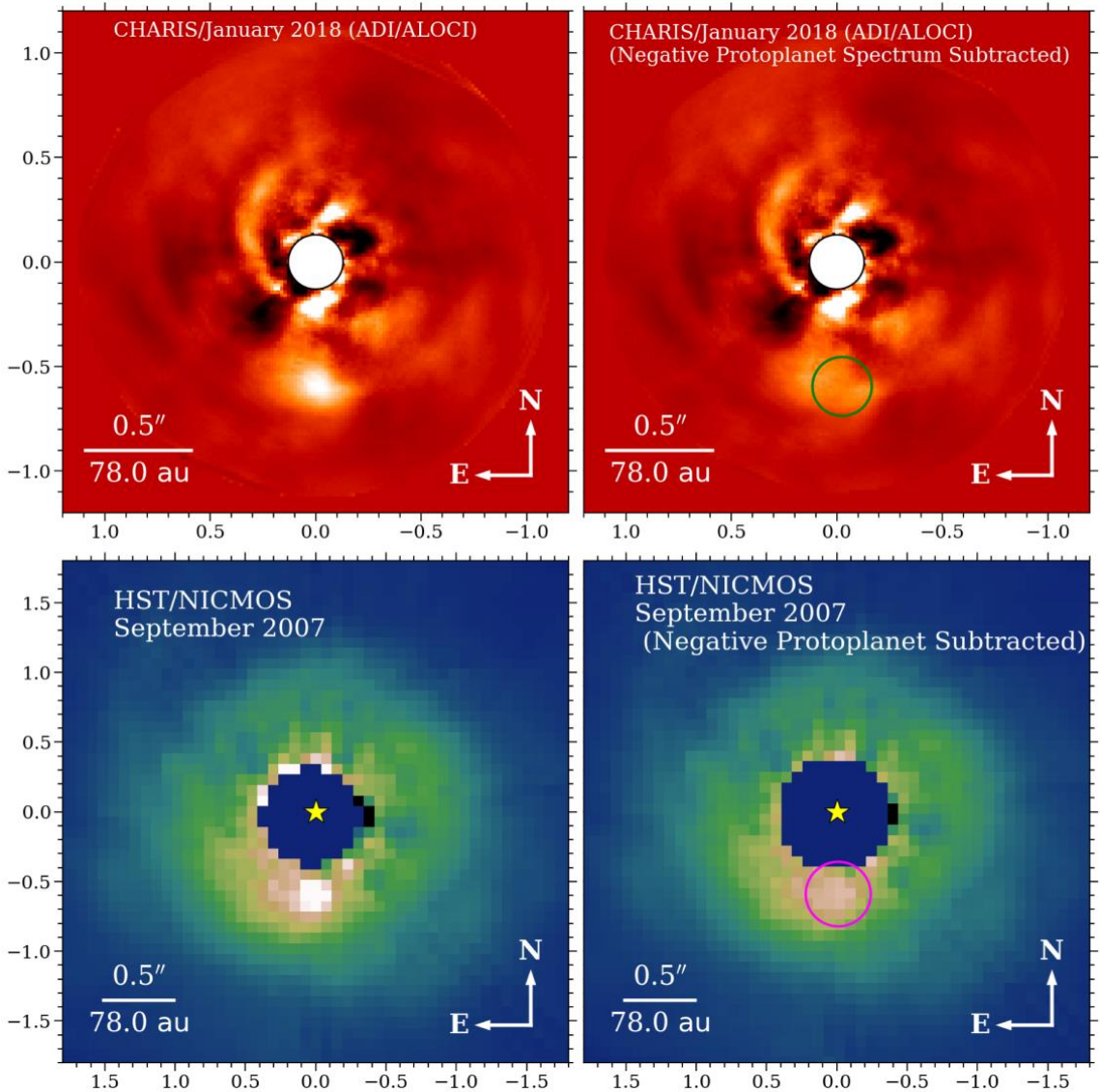

**Supplementary Figure 9 – Negative subtraction of AB Aur b.** (top panels) CHARIS ADI-reduced image of AB Aur compared to the same image where a negative copy of AB Aur b's spectrum with our assumed morphology is imputed into the observing sequence. (bottom panels) HST/NICMOS F110W PSF subtracted image compared to the same image with the model PSF for AB Aur b removed. In both cases, subtraction of the model protoplanet emission yields a flattened featureless background.

Absolute calibration of the CHARIS data draw from IRTF/SpeX data (PI M. Sitko). The SpeX data cover 0.8 to 5 microns: the measured flux densities agree with values for the optical spectrum . Multiple epochs of SpeX data for AB Aur show at most 10% variability in luminosity and no evidence for variability in the *shape* of the near-IR spectrum that could impact our results.

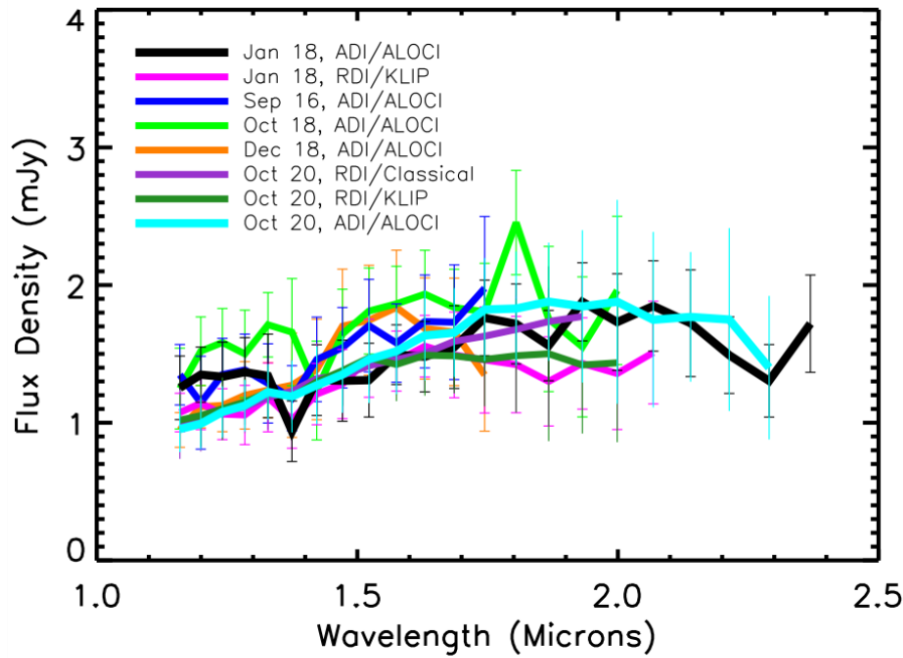

**Supplementary Figure 10 –- Spectra extracted for AB Aur b from different epochs and with different PSF subtraction algorithms.** The wavelength range covers only those channels where we can confidently claim a detection. The shapes of the spectra show excellent agreement across the J and H bands. Error bars denote 1-sigma uncertainties.

The spectrum extracted for AB Aur b is consistent across epochs, within errors. As shown in Supplementary Figure 10, in the J and H passbands where AB Aur b's detection is unambiguous, reductions using ADI and RDI, ALOCI and KLIP, yield consistent spectrophotometry. In Ks band, AB Aur b is generally not detected in the poorer quality September 2016, October 2018, and December 2018 data sets. However, spectrophotometry for the higher quality data agree very well within errors, regardless of reduction approach. Similarly, STIS photometry separately reduced from January and February 2021 agrees to within errors.

To assess whether AB Aur b can be explained by scattered starlight, we compare its spectrum with that from the protoplanetary disk extracted from multiple locations (Supplementary Figure 11). Regions include disk locations at positon angles flanking AB Aur b, disk locations at comparable angular separations but very different position angles, and extractions along the two spiral arms at smaller separations. In all cases, AB Aur b's spectrum differs. Supplementary Figure 12 (left panel) compares the normalized AB Aur b spectrum extracted from data processed with ADI/A-LOCI and the AB Aur protoplanetary disk processed with RDI/KLIP (both from January 2018 data). AB Aur b is substantially bluer than the AB Aur protoplanetary disk. Supplementary Figure 12 (right panel) shows the spectral energy distribution comparisons described in the main text. The accreting planet model includes contributions from a planet atmosphere and from magnetospheric accretion (see main text and Methods). As noted in the main text, a simple blackbody with a temperature comparable to that from the planet atmosphere model reproduces AB Aur b's IR photometry as well. The dissimilarity between the spectra of AB Aur b and the disk holds for different reduction approaches; each region of the disk for which we extracted spectra is redder than AB Aur b (Supplementary Figure 13).

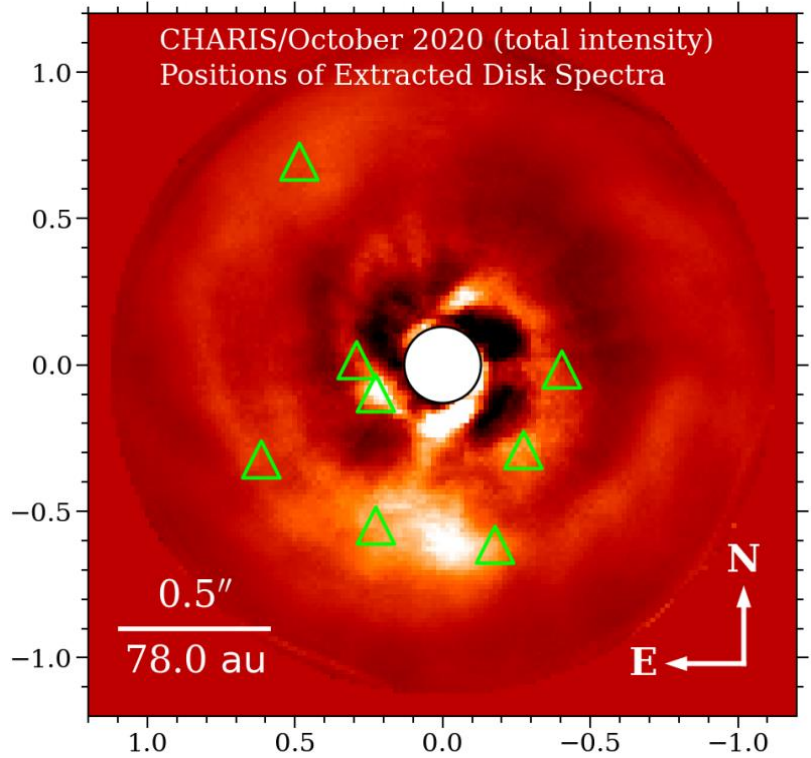

**Supplementary Figure 11– Locations of disk spectra extracted in Supplementary Figure 12** denoted by lime-green triangles.

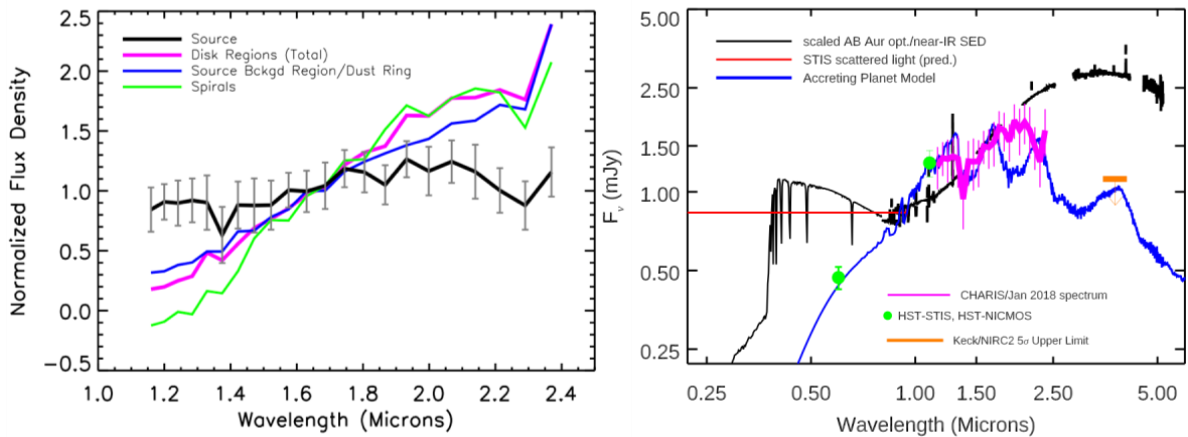

**Supplementary Figure 12 –SED modelling of AB Aur b.** (left) Normalized CHARIS spectrum of AB Aur compared to extracted spectra of different regions of the AB Aur disk. The AB Aur b spectrum derives from our ADI-ALOCI reduction of the January 2018 data; the disk spectra derive from the RDI-KLIP reduction of the same data. spectra extracted from individual locations and with different reductions show the same trends. (right) AB Aur data compared to a scaled spectrum of the primary and an accreting planet model. The model assumes a temperature of 2200K, gravity of log(g) = 3.5, and radius of 2.75 jovian radii. Magnetospheric accretion is truncated at 7.5 jovian radii; the planet mass is 9 jovian masses and accretion rate is $\dot{M} \sim 1.1 \times 10^{-6}$ $M_J$/yr. Other combinations of planet mass and accretion rate yielding $M\dot{M} \sim 1.1 \times 10^{-5}$ $M_J^2$/yr likewise can reproduce AB Aur b's SED; a more blackbody-like, featureless

spectrum matched to the near-IR data combined with magnetospheric accretion achieves comparable fits. Analysis in S5 considers the VAMPIRES H$\alpha$ and optical continuum detections. Error bars denote 1-sigma uncertainties.

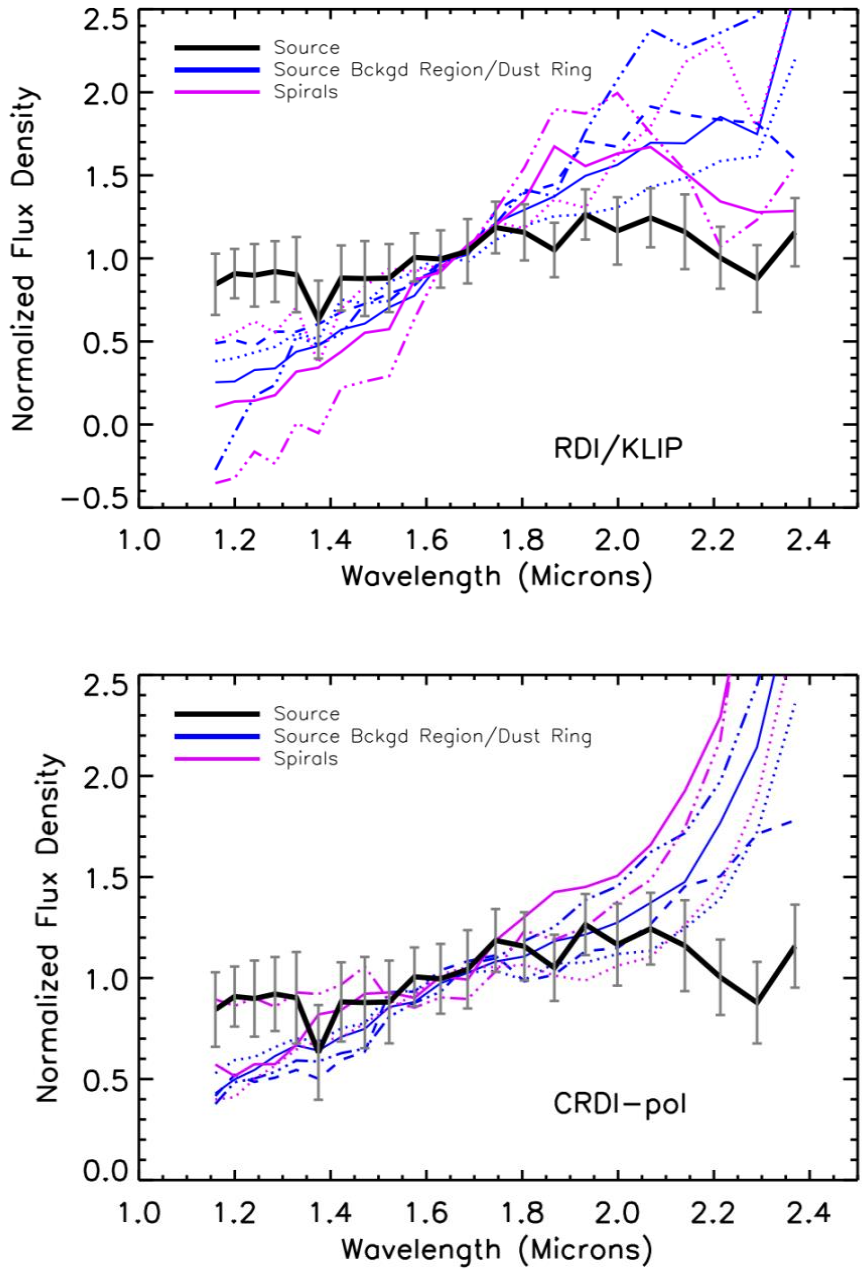

**Supplementary Figure 13 –Spectrum of AB Aur b compared to disk spectra extracted from locations depicted in Supplementary Figure 11.** The top panel shows our nominal comparison using RDI/KLIP-reduced data: the average of measurements for the dust ring and spirals is shown in Supplementary Figure 12, left panel. The bottom panel shows similar comparisons for data reduced using polarimetry-constrained RDI, which has poorer speckle suppression at small angular separations but higher throughput than the RDI/KLIP reduction. In both cases, the disk is redder than AB Aur b.

## S3. Astrometry of AB Aur b

Precise astrometry of AB Aur b first requires accurate absolute calibration of the ground-based instruments used for detection. The HST instrument astrometric calibrations are exceptionally robust: e.g. NICMOS's pixel scale and north position angle have a precision of $6\times10^{-3}$ mas and $2\times10^{-3}$ degrees, respectively, with a geometric distortion corrected to 0.18 pixels over the full detector[8]. The narrow field NIRC2 camera is precisely calibrated with a pixel scale and north PA offset of 9.971 mas/pixel and $0.262°$ with exceptionally small uncertainties of $4\times10^{-3}$ mas and $0.02°$, respectively[9].

Following our previous, preliminary analysis[6], we tie astrometric calibration of CHARIS to NIRC2 through epoch-matched observations of HD 1160 to obtain astrometry of the low-mass companion HD 1160 B. As a secondary astrometric calibration for CHARIS, we observed the M5 globular cluster, which has also been observed with HST/WFC3. For HiCIAO, we compared data for DH Tau and kappa And to NIRC2 obtained in December 2015/November 2016 and May 2016, respectively. We used astrometry of their companions to pin HiCIAO's pixel scale and north PA to NIRC2's.

CHARIS, HiCIAO, and NIRC2 data were processed following methods used for AB Aur. We used the IDL function cntrd.pro to estimate the centroid position of HD 1160 B, DH Tau B, kappa And b, and stars surrounding our M5 field. These analyses point to a slightly revised CHARIS pixel scale of 16.15 +/- 0.05 mas/pixel but the same north position angle offset as derived before: $2.20°$ +/- $0.27°$. We find no evidence for substantial differences in the north PA offset in CHARIS from epoch to epoch. For HICIAO, we find a pixel scale and north position angle of ~8.3 +/- 0.1 mas/pixel and -$1.0°$ +/- 0.1. At the separation of AB Aur b, these pixel scale and north PA uncertainties translate into positional uncertainties on the order of 3 mas: uncertainties due to the intrinsic SNR of the detection dominate over these systematic calibration uncertainties.

AB Aur b's computed position slightly varies with our initial guess for the position, typically ~0.1--0.3 pixels in both coordinates. Therefore, to derive robust astrometry for AB Aur b from ground-based data and estimate conservative errors, we constructed a grid of initial estimates of 4x4 pixels in size bracketing the apparent position and quantitatively determined the final centroid position and its uncertainty from the average and standard deviation of individual estimates drawn from background-subtracted images, weighted by the SNR at each estimated position. Since AB Aur b is not unambiguously detected in all spectral channels for every data set, we focused only on the channels with unambiguous detections. E.g. we included all channels for the 2018 January 06 and 2020 October 02 data but channels 1-15, 1-14, and 1-16 for the 2016 September 18, 2018 October 16, and 2018 December 21 data. To consider the north position angle uncertainty for CHARIS, we also included an astrometric uncertainty of 3 mas. Final errors include this centroid uncertainty estimate, the intrinsic SNR, and absolute calibration uncertainties for CHARIS and HiCIAO.

Table 1 in the main article describes the mean and 68% confidence intervals for AB Aur b's semimajor axis, eccentricity, and inclination for orbit fitting without restrictions on parameters. Supplementary Figure 14 shown below plots the 100 best-fitting orbits.

Having formed from a protoplanetary disk, AB Aur b may be coplanar with the disk. We therefore investigated a second set of fits adopting a Gaussian prior on the inclination centered on 25 degrees (+/- ~9 degrees). For this fit, the mean of posterior for the semimajor axis lies at ~94.19 au [80.22,108.14]; the eccentricity is a slightly lower 0.2 [0.08,0.32] while the inclination reflects the input prior. Other orbital parameters (e.g. longitude of the ascending node) remain poorly constrained.

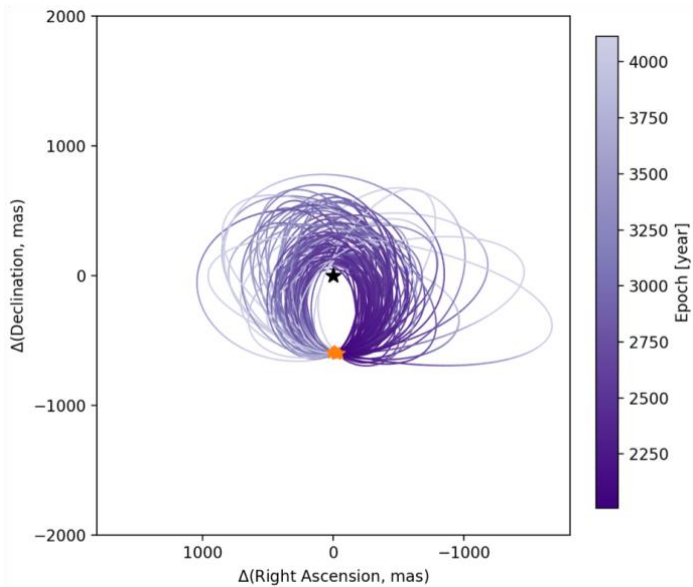

**Supplementary Figure 14 – The 100 best-fitting orbits for AB Aur b from our astrometric analysis.** Most solutions have AB Aur b near aphelion.

## S4. Wider-Separation Clump-like Signals

As described in the main text, HST data identify at least one, possibly two concentrated emission clumps at extremely large angular separations of $\rho \sim 2.75$" and 3.72" (429 au and 580 au): hereafter, sources "c" and "d". Supplementary Figure 15 displays these sources in the 1999 and 2021 STIS data (left, middle) and 2007 NICMOS data (right). In the STIS data, both sources are detected, with SNRs of 9—12 for "c" and 6—9 for "d". Only source "c" is visible in NICMOS data (SNR ~3). While we detect no other similar sources in the AB Aur disk, much of the north half of the AB Aur disk is blocked by the coronagraph in our most sensitive HST data set (2021 STIS data).

Following similar analysis for AB Aur b, we estimated the size of these two point-like features. Their radii are roughly ~0.063"—0.079" in size. Using the same methods adopted for AB Aur b, we measure STIS photometry of 20.40 +/- 0.11 and 22.48 +/- 0.15 magnitudes for "c" and "d", respectively. These two features at wider separation are not detected in polarimetry, although neither is the disk at such wide angular separations detected. These two features have STIS and NICMOS photometric measurements that can conceivably be matched by a scattered starlight (Supplementary Figure 16). Thus, they remain *candidate* sites of planet formation requiring further study.

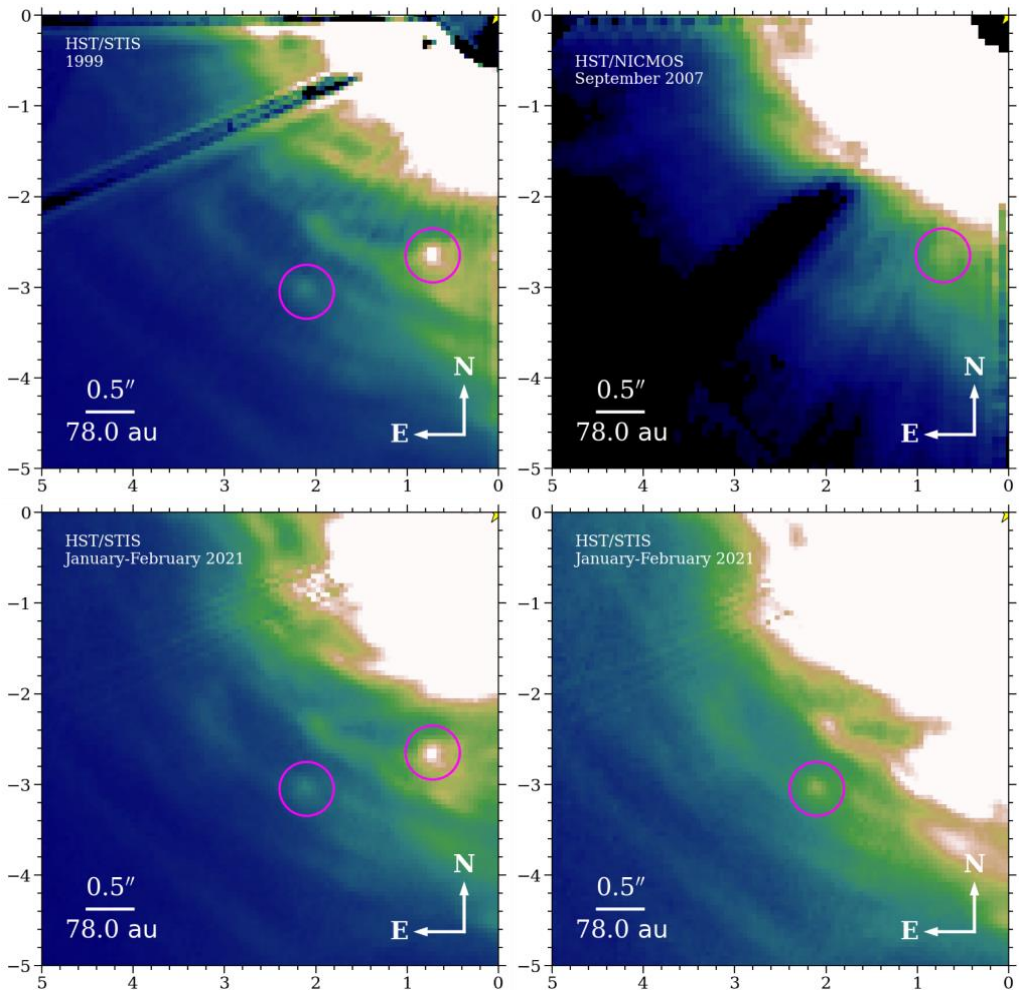

**Supplementary Figure 15 – Detection of two additional clump-like signals at ultra-wide separations in AB Aur's disk.** (top-left) HST/STIS data from 1999 showing the detection of sources 'c' and 'd' (brighter and dimmer source, respectively) at ~2.75" and ~3.72" and (top-right) HST/NICMOS data from 2007 showing a marginal detection of source 'c'. (bottom) Detection of source 'c' and 'd' in 2021 STIS data: the left panel is at a similar color stretch to that displayed for the 1999 STIS data.. The right panel has a more aggressive color stretch to more clearly show source 'd'.

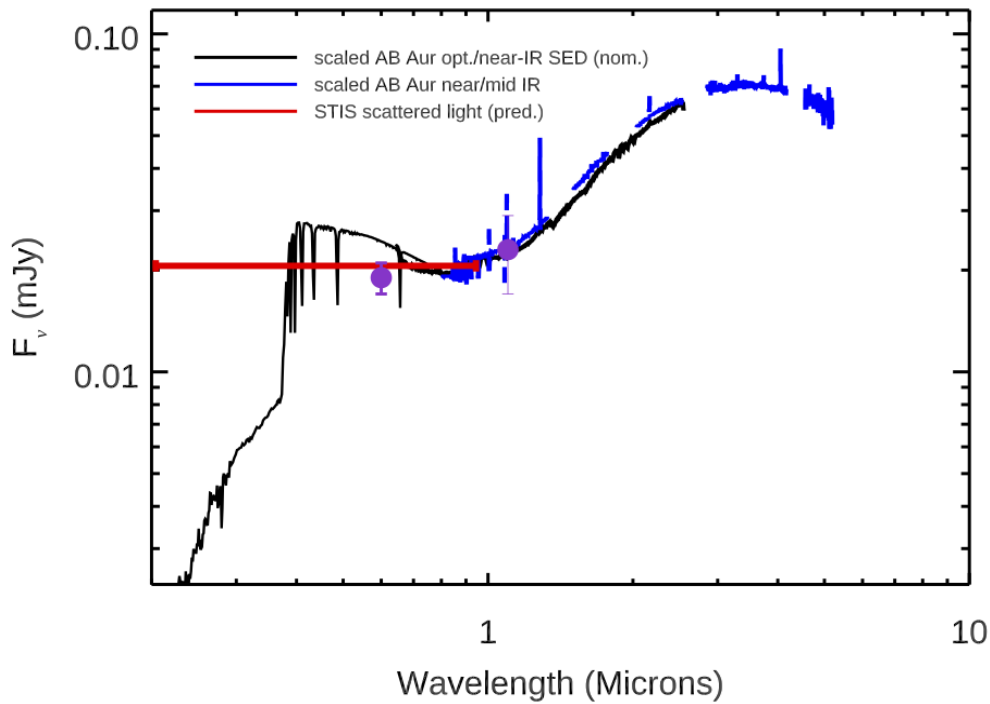

**Supplementary Figure 16 – Analysis of source 'c'.** Photometry of source c (purple dots) compared to the spectrum of AB Aur. Error bars denote 1-sigma uncertainties.

## S5. Details on Modeling AB Aur b's Optical Emission from VAMPIRES: Evidence for Accretion?

Supplementary Figure 12 (right panel) shows that the STIS, NICMOS, CHARIS, and NIRC2 data are inconsistent with a stellar photosphere + sub-au dust component expected for disk feature unrelated to a planet. For presentation clarity, we omitted comparisons to ground-based optical photometry from VAMPIRES. Here, we consider the VAMPIRES measurements in more detail.

AB Aur b's flux density in H$\alpha$ is nearly a factor of 3 higher than its continuum brightness (3.01 mJy vs 1.13 mJy or a factor of ~2.66). Normally, this high H$\alpha$ would be unambiguously interpreted as protoplanet accretion. However, AB Aur itself has strong H$\alpha$ from accretion: its line flux density is a ~2.4 higher than the continuum. As the H$\alpha$ detections themselves are only at SNR ~5—6, it is not possible to definitively tell whether H$\alpha$ is due to scattered starlight or accretion. Subtracting the continuum image from the H$\alpha$ image removes all of the protoplanetary disk scattered light on the differenced image (e.g. the spirals) but leaves residual emission at the ~2-$\sigma$ level at the approximate position of AB Aur b. Another challenge with interpretation is that VAMPIRES's astrometric calibration is still a work in progress: it is not clear whether VAMPIRES signal is entirely coincident with AB Aur b or is displaced in azimuth. Thus, we consider the VAMPIRES to be *consistent with* optical emission from AB Aur b but not strictly conclusive as is the case for CHARIS, STIS, and NICMOS data.

Future, deeper observations with VAMPIRES or higher-resolution spectra from other facilities (e.g. VLT/MUSE) may clarify whether this emission results from accretion or a scatererd light component.

## S6. Synthetic Scattered Light Images of AB Aur: An Embedded Protoplanet Model

We used the following approach to set up a Monte Carlo Radiative Transfer model of an embedded protoplanet in MCMax3D to reproduce the appearance of AB Aur b. The model adopts a temperature of 9770K and luminosity of 59 solar luminosities for the primary and an optical extinction of Av = 0.5 and a distance of 155.9 pc for the system. Scattered light imaging reveals the protoplanetary disk down to an inner working angle of 0.1" (~16 au) and exterior to ~600 au, where the majority of this light is within 300 au[10]. Millimeter imaging reveals emission concentrated near the star and in a ring approximately 0.75"—1.5" from the star (see main text). Thus, we considered two separate dust components: a population of small grains extending from the inner disk boundary out to 300 au and a more massive population of dust grains confined in a ring from 130 au to 250 au.

We varied properties of each dust population to simultaneously reproduce the surface brightness of AB Aur's disk in scattered light, its millimeter emission, and its spectral energy distribution. This parameter search led to the following values. For the first dust component, we set the inner edge of the disk to 0.27 au, similar to the 0.24 au value found from interferometric modelling[1], but with a rounded disk wall where the peak density is not reached until 0.5 au. The disk is flared with a scale height varying as $r^{1.3}$ with a shallow surface density power law of $r^{-0.5}$. Dust grains range from 0.1 μm to 1 μm, with a standard $\beta = -3.5$ power law size distribution, and a total dust mass of $1.75 \times 10^{-5}$ solar masses. For the second dust component, we set the inner edge of the disk to 130 au, similar to the deprojected radius of much of the disk seen in scattered light, again with a rounded disk wall where the peak density is not reached until 160 au. The disk is flared with the same scale height power law but has a steeper surface density power law of $r^{-1.0}$. Dust grains range from 1 μm to 100 μm, again with a standard $\beta = -3.5$ power law size distribution, and a total dust mass of $1.25 \times 10^{-4}$ solar masses. For both dust populations, we assume carbon-less grains with a 0.2 fractional porosity.

To generate a detailed model of the disk temperature structure and optical depth we used a Monte Carlo simulation with $10^7$ photons. Generating synthetic scattered light images used $5 \times 10^7$ photons, although we saw no discernable difference in the images after $10^7$ photons. Synthetic millimeter images required a comparable number of photons. While the disk model overpredicts the scattered light brightness at small (<0.3") separations, it accurately reproduces the surface brightness of the disk in H band at wider, >0.4" separations and reproduces the geometry of the dust ring seen in the millimeter.

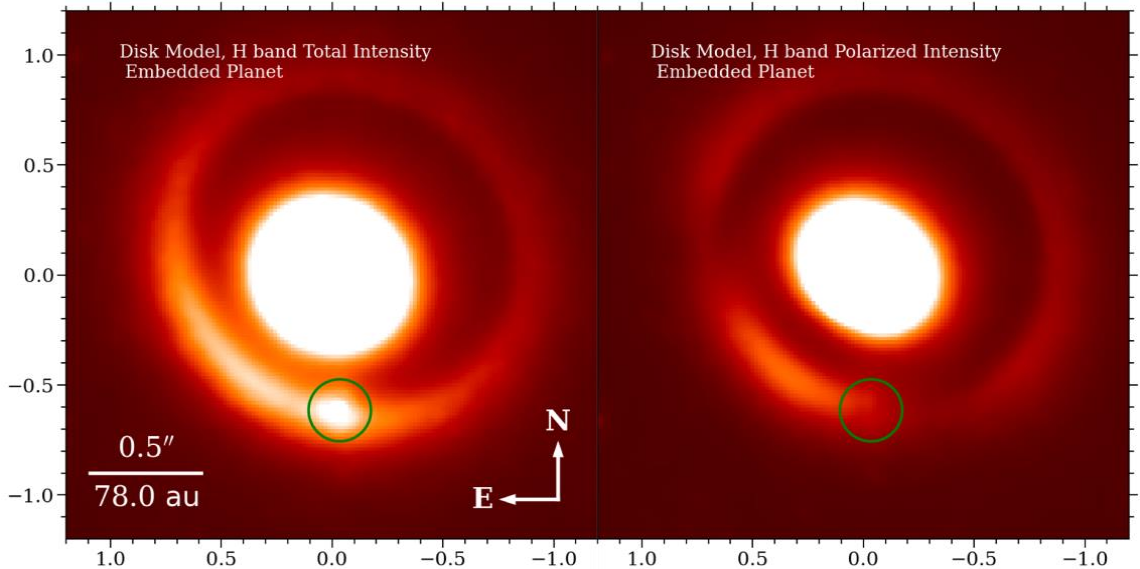

**Supplementary Figure 17 – Comparison between synthetic total and polarized intensity scattered light images for an embedded AB Aur b.** **(left)** Synthetic image of AB Aur, its protoplanetary disk, and an embedded AB Aur b at 1.63 microns in total intensity. (right) The same model in polarized intensity.

We estimated the optical depth along our line of sight at locations similar to AB Aur b by comparing its emission in absence of a disk to its emission embedded in a disk. At the companion's location, the disk is marginally optically thin/thick. The model shown in Figure 5 in the main text has an optical depth at AB Aur b's location of $\tau \sim 1.3$ in H band along our line of sight, although models with slight parameter variations of $\tau = 0.25$—2 also reproduce the scattered light brightness of the disk. In these conditions, AB Aur b emits light which is scattered off nearby dust grains, elevating the surface brightness of the image over a radius of ~0.05" in the simulated image. When convolved with the CHARIS PSF, AB Aur b appears spatially extended, with a FWHM of ~0.1", similar to the measured size of the companion in CHARIS. The signal from AB Aur b over this larger area is roughly equal to its point-source emission in absence of any intervening protoplanetary disk material. For models with progressively higher optical depths ($\tau \gg 10$), the near-IR emission becomes extremely diffuse with less measurable signal and eventually invisible. Similarly, for our nominal model the emission at optical wavelengths is more diffuse.

While this model is meant purely as a proof of concept and not a detailed, quantitative fit to the data. We further emphasis that this model is agnostic as to the underlying mechanism responsible for a planet being located at wide separations: core accretion or disk instability.

Importantly, however, the model does reproduce key aspects of the CHARIS images. It does predict that AB Aur b should appear extended. While the overall polarization level of the synthetic model is low compared to observations, the synthetic polarized intensity image does not reveal any elevated emission at the position of AB Aur b (Supplementary Figure 17). Critically, the detection of some polarized emission at AB Aur b's position in the model and in the real data does not mean that AB Aur b itself is detected in polarized light because AB Aur b is embedded in a disk which should have a polarized light signal. Rather, the lack of concentrated emission at the location of AB Aur b in polarized light implies that it is not detected.

In order to generate an appearance of AB Aur b consistent with our data, the model requires a narrow range of optical depths: marginally optically thin values for the disk optical depth at the protoplanet's position. However, the surface brightness of the disk is consistent with observed CHARIS values. Within the framework of MCMax3D, the disk model parameters chosen are then those that reproduce the near IR scattered light images irrespective of whether or not there is a planet.

## S7. Tying AB Aur Images to Simulations of Planet Formation: A Disk Instability-Produced Clump

Details for how we construct a model of planet formation by disk instability are as follows. We construct a model two-dimensional disk in cylindrical coordinates. We then solve for the vertically integrated equations of motion with self-gravity and finite cooling

$$\frac{D\Sigma}{Dt} = -\Sigma \nabla \cdot \boldsymbol{u}$$
$$\frac{D\boldsymbol{u}}{Dt} = -\frac{1}{\Sigma}\nabla P - \nabla \Phi + \frac{1}{\Sigma}\nabla \cdot \boldsymbol{\zeta}$$
$$T\frac{Ds}{Dt} = -c_V \frac{(T - T_{\text{ref}})}{\tau} + \Gamma_{\text{sh}}$$
$$P = \Sigma c_s^2/\gamma$$
$$\Phi = \Phi_{\text{sg}} - \frac{GM}{r}$$
$$\nabla^2 \Phi_{\text{sg}} = 4\pi G\Sigma \delta(z)$$

where $\Sigma$ is the density, $\boldsymbol{u}$ is the velocity, $P$ is the pressure, $\Phi$ is the gravitational potential, $\zeta$ is the shock viscosity tensor, $T$ is the temperature, $s$ is the entropy, $c_v$ is the specific heat at constant volume, $T_{\text{ref}}$ is a reference temperature, $\tau$ is the thermal time, $\Gamma_{\text{sh}}$ is the shock heating, $c_s$ is the sound speed, $\gamma$ is the adiabatic index, $\Phi_{\text{sg}}$ is the disk selfgravity, $G$ is the gravitational constant, $M$ is the stellar mass, and $r$ the stellocentric distance. The advective derivative is

$$\frac{D}{Dt} = \frac{\partial}{\partial t} + \boldsymbol{u} \cdot \nabla$$

We use the Pencil Code[11] to solve the equations of motion. We capture shocks using an explicit shock viscosity prescription. The third term in the momentum equation is the viscosity required to spread shocks out to resolvable width, and the last term in the entropy equation is the shock viscous heating. The shock viscosity tensor takes the form of a bulk viscosity

$$\zeta_{ij} = v_{\text{sh}}\Sigma\delta_{ij}\nabla \cdot \boldsymbol{u}$$

and the associated shock heating is

$$\Gamma_{sh} = v_{sh}(\nabla \cdot \boldsymbol{u})^2$$

They depend on the shock viscosity, which we define numerically as

$$v_{sh} = c_{sh} \langle \max_5[(-\nabla \cdot \boldsymbol{u})^+]\rangle \min(\Delta x)^2$$

The actual form of the shock viscosity that we use has been described in [12]. The superscript plus sign indicates the positive part of the quantity. The "5" subscript indicates the maximum is taken within 5 grid zones. As long as the shock is resolved, the value of the shock viscosity coefficient does not change the amount of heating; rather, it just changes the volume (number of grid cells) over which the shock energy is spread.

The reference temperature $T_{ref}$ is set to the initial temperature at every radius. Sixth-order hyper-dissipation terms are added to the evolution equations to provide extra dissipation near the grid scale, as discussed in [13]. These terms are needed for numerical stability because the high-order scheme of the Pencil Code has little overall numerical dissipation[14]. They are chosen to produce Reynolds numbers of order unity at the grid scale, but then drop as the sixth power of the scale at larger scales, so that they have negligible influence on the large-scale flow.

The disk ranges from 30 to 300 au, and full $2\pi$ azimuthal coverage. The resolution is 432 grid cells in radius, and 864 cells in azimuth. The grid is logarithmically spaced in radius, and linear in azimuth.

For boundary conditions, we use sponge zones in the radial boundaries, that drive all quantities back to the initial condition. The width of the sponge is 3 au at the inner boundary and 30 au in the outer boundary. The quantities are driven back to the initial condition within $10^{-2}$ orbit at 100 au. Boundary condition for radial velocity is zero gradient; azimuthal velocity, density, and entropy use constant gradient.

The density is initially set as an exponentially truncated disk

$$\Sigma = \Sigma_0 \exp\left[-\left(\frac{r}{r_0}\right)^2\right]$$

with $r_0$=100 au. In the inner disk (30-70 au), this density profile is well approximated by a power-law falling as square root of the radius. The temperature follows a power law

$$T = T_0 \left(\frac{r}{r_0}\right)^{-q}$$

with $q$=1. $T_0$ is set to 30K at 100 au. The sound speed is given by

$$T = \frac{c_s^2}{c_p(\gamma - 1)}$$

where $c_p = \gamma c_V$ is the specific heat at constant pressure. The specific heats are related to the universal gas constant $R$ by

$$R = \mu(c_p - c_V)$$

where $\mu$ is the mean molecular weight of the gas, set to 2.3, for a 5:2 $H_2$-He mixture. Given $\gamma=1.4$, the sound speed at 100 AU is about 0.33 km/s. This sets the disk scale height $H=c_s/\Omega_K$ (where $\Omega_K$ is the Keplerian frequency) at 100 au as 0.07, considering a central mass of $M=2.4$ solar masses.

The reference volume density $\rho_0$ is $10^{-12}$ g/cm³, leading to a column density of $\Sigma = 2.7$ g/cm² at 100 au. These physical conditions translate into Toomre $Q$ values

$$Q = \frac{c_s \Omega}{\pi G \Sigma}$$

below 1 between 50 and 150 au.

The initial radial velocity is zero; the initial azimuthal velocity is corrected by the pressure gradient and self-gravity

$$\dot{\phi}^2 = \Omega_K^2 + \frac{1}{r}\left[\frac{1}{\Sigma}\frac{\partial P}{\partial r} + \frac{\partial \Phi_{sg}}{\partial r}\right]$$

noise added in the velocity field at the 5 x $10^{-3}$ $c_s$ level. The cooling time is set to $\beta = \Omega\tau$ =10 orbits. The gas self-gravity is solved via fast Fourier transforms in cylindrical coordinates with the method of logarithmic spirals[15]. Both the coefficients of shock and hyper diffusivity are set to 20.

The Bonnor-Ebert mass, which gives the collapsing mass for an isothermal sphere in a medium of external pressure $p0$ is

$$M_{BE}(p_0) = \frac{225}{32\sqrt{5\pi}} \frac{c_s^4}{(aG)^{3/2}} \frac{1}{\sqrt{p_0}}$$

where $a=1$ for constant density, and $a=1.67$ for a sphere that is denser in the center. For our choice of parameters ($\rho = 10^{-12}$ g/cm³, T=30K, $\gamma=1.4$), the Bonnor-Ebert mass is 0.93 Jupiter masses, which sets the expectation for the mass of the clumps formed in the model.

Within a few orbits, the disk breaks into fragments (Supplementary Figure 18). The left panels show the gas density, the right panel the local disk scale height considering self-gravity[16]

$$H' = \frac{H_{sg}}{2}\left(\frac{H_*}{H_{sg}}\right)^2 \left[\sqrt{1 + 4\left(\frac{H_{sg}}{H_*}\right)^2} - 1\right]$$

where $H_{sg} = QH$ and $H_* = (\pi/2)^{1/2}H$. Notice that the location of the clumps corresponds to regions of *low* scale height, evidencing $Q \ll 1$.

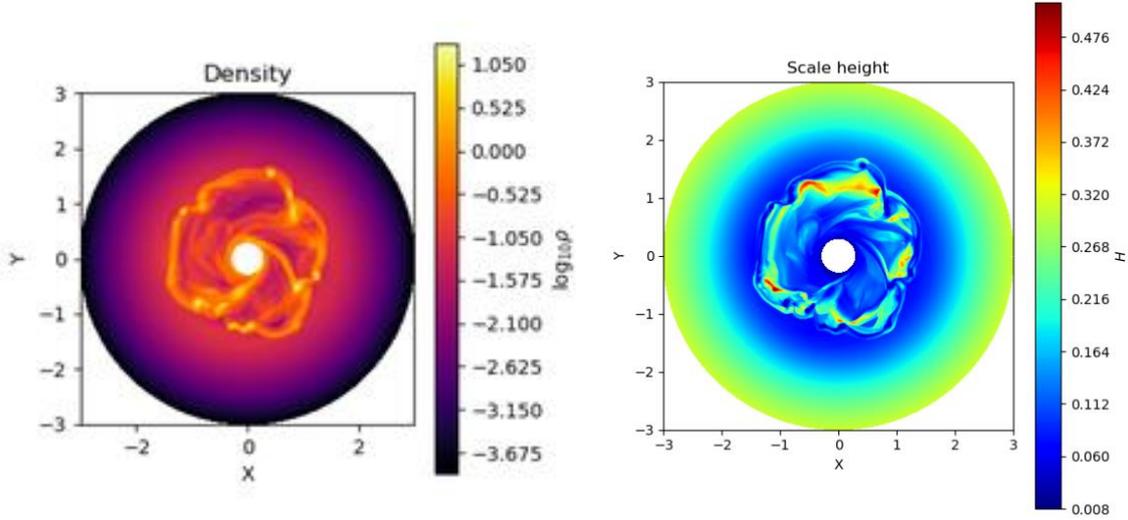

**Supplementary Figure 18** The state of the disk after 10 orbits. The unit of the axes is 100 au. *Left:* Density is shown in units of $10^{-12}$ g/cm$^3$. The disk has fragmented into many gravitationally unstable clumps. *Right*: The local scale height. The dark region just counter-clockwise from the scale height peak at the 8 o'clock position, identifies the collapsing clump's position. Other clumps have lower density and are cooler and thus are harder to see in scattered light.

To produce a synthetic image, we perform full radiative transfer post-processing with RADMC-3D[17], a popular Monte Carlo radiative transfer software. RADMC-3D combines a Monte Carlo code with a ray-tracing mode to simulate observations of the resulting temperature distribution. In practice, we first use thermal Monte Carlo simulations to determine the temperature of the optically thin regions of the disk. Then, a ray-tracing computation is used to create a synthetic image of what would be observed at infrared wavelengths of interest. The last snapshot was taken from the hydrodynamical simulation and its parameters were input into RADMC-3D, using the pipeline developed in [18].

Since the hydrodynamical model is 2D, we extend the model in the vertical direction using the scale height shown in Figure 1, and assuming hydrostatic equilibrium

$$\rho = \rho_0 \exp\left[-\frac{z^2}{2H^2}\right]$$

The grid is extended in spherical coordinates, with 128 points in the meridional direction, covering 10 scale heights above and below the midplane.

The central star is 2.5 solar radii, at effective temperature 9770 K. We assume well-coupled micrometer-sized dust grains to be perfectly coupled to the gas. The dust density then follows directly from the Pencil Code output, scaled down by a factor of 100 (the dust-to-gas ratio) and converted into cgs units of grams per cubic centimeter. We use the wavelength-dependent opacities from [19]. The radiative field is sampled with $10^7$ photons, out of which are treated as scattering. To reproduce AB Aur b, an extra source of photons is placed at (X,Y)=(-90,-80) au. The radius of the source is set to 2 Jupiter radii, with temperature 5000 K. While some studies find far cooler temperatures associated with GI[20-21], they typically assume an isothermal gas disk or do not include accretion luminosity of the contracting clumps with the central gravitational potential not well resolved. In contrast, other approaches find central clump temperatures of at least 1500K, up to 5000--6000 K[22-24].

As with the scattered light model of an embedded planet, this disk instability model is a proof-of-concept and is not intended to be a rigorous approach to precisely reproduce all aspects of the observed IR data. Among its potential weaknesses, the disk instability model produces many clumps, not one, although these clumps are of different densities or at least different locations such that they are not all equally visible. The detectability of only one may require unique conditions in the disk and/or viewing geometry effects.

Nevertheless, the model does capture key properties of our IR images. Namely, it produces a bright clump-like signal at $\sim 10^{-4}$ contrast at an angular separation of 0.6" (~115 au, deprojected): consistent with our data. Furthermore, it produces spiral density waves, where the signal we interpret as AB Aur b lies at the terminus: consistent with the modeling of spirals seen in ALMA data.

Taking together these simulations and the observed system properties lends strong credibility to the view that AB Aur b is protoplanet being formed by disk instability. Other recent work likewise identifies the AB Aur system as a potential site for planet formation by disk instability[25], although they focus on features possibly interpreted as planets at smaller separations based on VLT/SPHERE results (see next Section).

Regardless of the details of AB Aur b's exact formation history, AB Aur b identifies evidence for a planet formation on wide separations and the system shows clear signs of having gravitationally unstable disk, which likely influence its formation. Future observations from the ground and space as well as future, sophisticated numerical simulations may clarify which interpretation is better explains the system.

### S8. The Emission Source for AB Aur b: Planet Atmosphere, Circumplanetary Material, or Both?

While analysis in S2 shows that AB Aur b is not scattered starlight misinterpreted as a protoplanet, pinpointing the object's emission source is challenging. The CHARIS and NICMOS infrared detections and NIRC2 upper limits are consistent with thermal emission in the ~2200 K range. However, emission this cold cannot explain the STIS photometry. Thus, in S2 we considered a composite model, consisting of a ~2200 K planet atmosphere to fit the IR data and a much hotter component tracing magnetospheric accretion to fit our optical data. We emphasize, though, that this fit is not unique: i.e. just because the planet atmosphere model reproduces the shape of the IR spectrum, it does not follow that we *must* be seeing the photosphere of a planetary-sized object. Blackbody emission of a comparable temperature – such as one might expect from a circumplanetary material --likewise can reproduce AB Aur b's IR spectrum.

Thus, AB Aur b's emission is consistent with two possibilities: 1) emission solely from circumplanetary material (i.e. a circumplanetary disk or envelope) or 2) emission from a ~2200 K planetary atmosphere with a roughly Jupiter size scale *and* a hotter circumplanetary component. The data themselves do not yet allow us to easily distinguish between these two scenarios. However, Scenario 1) is likely favored because embedded protoplanets are expected to be surrounded by either flattened circumplanetary disks or spheroidal circumplanetary envelopes, both of which are fractions of a Hill sphere in size and obscure any nascent planet atmosphere[26]. The hot emission component would likely originate from the shockfront of a circumplanetary disk/envelope instead of a planet atmosphere. For a roughly spherical circumplanetary envelope, radiative transfer resembles that of a stellar atmosphere. Photons from the embedded planet are successively absorbed, scattered, and re-emitted until they escape. The envelope/disk could contain well mixed dust grains whose small, micron-to-submicron sizes resemble those typical of protoplanetary disks. In such a case, the emission spectrum would be relatively featureless, lacking the sawtooth-like IR spectrum shaped by water, methane, and carbon monoxide opacity as expected for a ~2200 K photosphere[27].

Follow-up data spanning a wider range of wavelengths may clarify AB Aur b's emission source(s). If some of AB Aur b's IR emission originates from a planetary atmosphere, higher resolution and higher quality data may reveal expected molecular absorption. Better sampled red optical photometry may constrain the temperature of the hot emission component. AB Aur b's non-detection at Lp strongly disfavors at least some models of an extended, accreting circumplanetary disk that intercepts and reprocessed emission[22]. However, a future mid IR detection of AB Aur b drawn from higher-quality data may provide evidence of cooler circumplanetary disk emission and motivate new models of circumplanetary disks, circumplanetary envelopes, and/or accretion-driven luminosity.

## S9. Comparison with VLT/SPHERE Near-Infrared Results from Boccaletti et al. (2020) and Other Recent Work

As described in S1, we recover AB Aur b through an independent reduction of VLT/SPHERE total intensity data first published in [28]. The authors graciously allowed us to visually inspect their total intensity reduction. Modest field rotation does cause the ADI-reduced total intensity data to be heavily self-subtracted. However, upon a joint reinspection with the authors, AB Aur b is visible in their data for extremely conservative reductions, albeit attenuated by processing (A. Boccaletti, pvt. comm.), consistent with our detections. Greater field rotation and the ability to use a reference star for PSF subtraction allowed us analyze the same system from CHARIS data in a way that is less affected by algorithmic biasing.

The SPHERE paper identified other structures in the AB Aur disk potentially connected to planet formation. In addition to resolving spiral structure down to ~0.1", the study identified other structures in the disk possibly connected to planets that are not the focus of this work. Specifically, they identified a twist in one of the spiral arms at $\rho$ ~ 0.16" at a position angle of ~203.9° ("f1") and a point-like clump at $\rho$ ~ 0.68", 7.8° qualitatively similar to sources c and d in this work.

We easily resolve the inner spirals in most CHARIS data sets; the high-quality January 2018 and October 2020 data processed using reference star subtraction provide a particularly clear view of this region minimally biased by processing. Inspecting the region at the "f1" position does not unambiguously reveal a point source although it may identify a local brightness maximum in the spiral. Between the two data sets, the intensity of the spiral region at "f1" is the same to within 20%, in contrast to the order of magnitude difference identified by Boccaletti et al. and attributed to signal-loss from ADI. We do not find clear evidence yet that the spiral related to "f1" is rotating counterclockwise, although a longer time baseline and PDI data more amenable to removing the stellar halo may reveal such motion.

While our work was under final review, a Large Binocular Telescope-led (LBT) group posted preprints for two studies – Betti et al. [29] and Jorquera et al. [30] -- using 2—4 μm imaging to assess properties of the AB Aur disk and identify any companions. Over this wavelength range, they found that AB Aur's disk has a very red scattered-light disk color. They do not identify any companions, focusing their analyses on the non-detection of candidate sources identified from SPHERE polarimetry.

Visual inspection shows that our SCExAO/CHARIS data yields a far sharper view of the AB Aur system than the LBT data sets and thus should be viewed as authoritative over the latter. The LBT team does not detect spirals in their shortest wavelength (Ks band) data, attributing the non-detection to optical depth effects or variability. In comparison, the spirals are detected in every SCExAO/CHARIS and HiCIAO dataset shown in Figure S1, obtained over the course of 2 years (e.g. see Figure S3, right panels). Our data's advantages are likely due to the higher-order adaptive optics correction enabled by SCExAO coupled with more high-contrast imaging friendly seeing experienced on Maunakea. We also note that our detection is at lowest contrast with the disk at Ks: imaging AB Aur b at longer wavelengths will be even more challenging.

Both our Keck/NIRC2 Lp data and the LBT Lp data resolve the ring and detect the inner spirals but do not detect AB Aur b. The LBT data report contrast limits by calculating pixel-to-pixel noise maps from the standard deviation of values within a sliding box of 1.5 $\lambda/D$. The noise map for a point source covering many pixels, though, should be calculated from the robust standard deviation of pixels in a convolved image (e.g. convolved with a top-hat filter sized to a point source or instrumental PSF), which is typically far larger than the per-pixel standard deviation. Detection limits for an extended source like AB Aur b will also be different from that of a point source. Future imaging with facilities like the upgraded Keck/NIRC2 or JWST will improve upon our limits for AB Aur b's emission in the thermal IR.

The spiral structure within 1 arcsecond of AB Aur and the morphology of the disk at wider separations will be the subject of future CHARIS and HST studies.

## S10. AB Aur in the Context of other Directly Imaged Protoplanetary Systems

AB Aur now joins PDS 70 as a system with directly-imaged protoplanets. Like AB Aur b, PDS 70 bc show orbital motion, near IR photometry inconsistent with scattered light from the star, and a low polarization consistent with its emission originating from a planet, not a pure disk feature[29]. AB Aur b and PDS 70 bc also show H$\alpha$ emission[4], although AB Aur b's detection may have origins other than accretion. In contrast to PDS 70 bc, though, AB Aur probes embedded planet formation at ~100 au around massive stars, not the protoplanets lying within a cleared disk and orbiting on far smaller, solar system scales around stars less massive than the Sun. Thus, the two systems are laboratories for understanding planet formation on different scales, at evolutionary states, around different kinds of stars, and likely due to different formation mechanisms.

Other systems may harbour candidate protoplanets. For instance, two studies presented detections of protoplanets around the young, Sun-like star LkCa 15 from sparse aperture masking interferometry (SAM) and H$\alpha$ differential imaging[30-31]. However, later observations showed that the SAM detections correspond to disk features[32] and the H$\alpha$ emission may also originate from the disk[33]. Being a young, solar-mass star with candidate companions at ~20 au, the system also bears more of a resemblance to PDS 70 than to AB Aur.

HD 100546 is a 2 solar-mass star with one protoplanet candidate at ~50 au (HD 100546 b) and another at ~13 au (HD 100546 c), just interior to the gap in the protoplanetary disk[34-36]. The system is more similar to AB Aur than PDS 70 or LkCa 15. However, the nature of the candidates remains tentative. Ambiguities in the astrometry and spectra of HD 100546 b complicate its interpretation[37]. HD 100546 c is thus far reported in a single epoch in the peer-reviewed literature: follow-up in the near term may be precluded if the candidate is passing behind the disk wall[38]. Similarly, several studies report the detection of candidate companions around MWC 758 in the thermal IR: one on a solar system scale and one at a wide separation at the terminus of a spiral arm[39-40]. However, follow-up observations have yet to confirm these companions to establish orbital motion and clearly distinguish them from disk features[40-41]. Lacking the orbital motion, disambiguation from protoplanetary disk scattered light, and spectral confirmation obtained with AB Aur b, HD 100546 bc and MWC 758 bc remain candidates requiring further study and confirmation.